\documentclass[useAMS,usenatbib]{mn2e}
\usepackage{graphicx}
\usepackage{amssymb}

\newcommand{\apj}{ApJ}

\newcommand{\aap}{A\&A}

\newcommand{\MC}{\multicolumn}
\newcommand{\kms}{km\,s$^{-1}$}

\newcommand{\HII}{H{\sc ii}}
\newcommand{\sunn}{$_{\odot}$}
\newcounter{qub}

\DeclareRobustCommand{\ion}[2]{%
\relax\ifmmode
\ifx\testbx\f
{\mathrm{#1\,\textsc{#2}}}\else
{\mathrm{#1\,\mathsc{#2}}}\fi
\else\textup{#1\,{\mdseries\textsc{#2}}}%
\fi}

\title[Monitoring of DDO68 'Northern Ring' during 2016 to 2023]
{Monitoring of DDO68 'Northern Ring' SF regions during years 2016--2023}

\author[S.A. Pustilnik, Y.A. Perepelitsyna, A.S.~Vinokurov et al.]
{S.A.~Pustilnik,$^1$\thanks{sap@sao.ru (SAP)} Y.A.~Perepelitsyna,$^1$ A.S.~Vinokurov,$^1$  E.S.~Egorova,$^{2,3}$
A.S.~Moskvitin,$^1$
\newauthor V.P.~Goranskij,$^2$  A.N.~Burenkov,$^1$ O.A.~Maslennikova,$^1$ O.I.~Spiridonova$^1$ \\
$^1$ Special Astrophysical Observatory of RAS, Nizhnij Arkhyz, Karachai-Circassia 369167, Russia \\
$^2$ Sternberg Astronomical Institute, Lomonosov Moscow State University, Universitetskij Pr. 13,  Moscow 119992, Russia\\
$^3$ Astronomisches Rechen-Institut, Zentrum f\"{u}r Astronomie der Universit\"{a}t Heidelberg, M\"{o}nchhofstra\ss e 12-14,
     69120 Heidelberg, Germany}

\begin{document}

\label{firstpage}

\date{Accepted 2024 October 4, Received 2024 April~20}

\pagerange{\pageref{firstpage}--\pageref{lastpage}} \pubyear{2024}

\maketitle

\begin{abstract}

DDO68 is a star-forming (SF) dwarf galaxy residing in a nearby void. Its gas metallicity is among
the lowest known in the local Universe, with parameter 12+log(O/H) in the range of 6.96--7.3 dex.
Six of its SF regions are located in or near the so-called 'Northern Ring', in which the Hubble Space
Telescope (HST) images reveal many luminous young stars. We present for these SF regions (Knots) the
results of optical monitoring in 35 epochs during the years 2016--2023. The data was acquired with the
6m (BTA) and the 1m telescopes of the Special Astrophysical Observatory and the 2.5m telescope of the MSU
Caucasian Mountain Observatory. We complement the above results with the archive data from 10 other telescopes
for 11 epochs during the years 1988--2013 and with 3 our BTA observations between 2005 and 2015.
Our goal is to search for variability of these Knots and to relate it to the probable light variations of
their brightest stars. One of them, DDO68-V1 (in Knot~3), was identified in 2008 with a luminous blue variable
(LBV) star, born in the lowest metallicity environments. For Knot~3, variations of its integrated light in the previous
epochs reached $\sim$0.8~mag. In the period since 2016, the amplitude of variations of Knot~3 reached $\sim$0.3~mag.
For the rest Knots, due to the lower amplitudes,  the manifestation of variability is less pronounced.
We examine the presence of variability via the criterion $\chi^{2}$ and the Robust Median Statistics
and discuss the robustness of the detected variations.
The variability is detected according to the both criteria in the lightcurves of all Knots
with the $\chi^{2}$ confidence level of $\alpha$ = 0.0005.
The peak-to-peak amplitudes of variations are $\sim$0.09, $\sim$0.13, $\sim$0.11, $\sim$0.08 and $\sim$0.16~mag for
Knots~1, 2, 4, 5 and 6,  respectively. The amplitudes of the related variations of the brightest supergiants in these
regions can reach of $\sim$3.0~mag.
\end{abstract}


\section[]{INTRODUCTION}
\label{sec:intro}

Massive stars (conditionally, of 8 to 100 and more M\sunn) are very important element of
several interrelated directions in astrophysics, from star formation and its feedback, galaxy formation
and evolution to cosmological issues related to reionisation of the intergalactic medium.
Due to their short life timescale, the massive stars in the Galaxy have metallicities close to the current gas
metallicity, that is close to Z$\sim$Z\sunn.

There are several tasks related to the study of massive stars in the context of the diverse metallicity.
In particular, the understanding of properties of massive stars with very low Z of $\lesssim$Z\sunn/30 is crucial
for studying and modelling of galaxy formation and evolution at the epoch of dawn of the Universe.
The need to check the modern models of massive very low-Z stars on the real objects assumes
the search for and study such objects  in the outer galaxies.

The gas metallicity in the nearby Universe was found to vary in the range of ($\sim$0.02--3)~Z\sunn.
Due to the well-known relation between  mass (or luminosity) and gas metallicity in late-type galaxies,
the lowest metallicity massive stars are expected to form currently in dwarf galaxies.
According to the statistical relation between the gas metallicity and galaxy luminosity
for late-type galaxies within the Local Volume (LV) \citep{Berg12}, dwarf galaxies
with the gas metallicity (and hence, that of young massive stars) of Z $\lesssim$ Z\sunn/30 (eXtremely Metal-Poor,
hereafter, XMP) should
be rather faint, with M$_{\rm B} \gtrsim$ --9~mag. Such faint dwarfs and their massive stars are accessible only
at the distances of the LV and its environs. The nearest dwarf with that low metallicity, Leo~P \citep{Skillman13},
possesses the only \HII\ region, excited by one O7-8~V star. It was recently studied by \citet{Telford23}. To substantially
advance in the studying of massive stars at such extremely low Z (including the massive stars in the evolved
stages), one needs XMP galaxies with many massive stars within the LV \citep{Garcia21}.

According to our spectral study of void galaxies \citep{PaperVII, XMP.SALT, XMP.BTA, SALT2},
the XMP dwarfs favour the void environment. Moreover, for a given luminosity, they show substantially reduced
gas metallicities. And visa versa, for their extremely low metallicities, they appear substantially more luminous
than one would expect from the reference relation of  \citet{Berg12}, derived for galaxies in the denser environment.
Therefore, statistically, the probability to find multiple massive stars
in void XMP dwarfs is higher than for the similar dwarfs outside voids.

Luminous blue variable (LBV) stars are thought to represent a relatively  short
transient and rather unstable stage of massive star evolution from the main
sequence hydrogen burning O-stars to the core-helium burning Wolf-Raye (WR)
stars \citep{HD94}.
During this stage, massive stars lose the substantial mass via the powerful
wind and in a series of 'normal' eruptions with the typical timescales
of a few years per event. Besides, some evidence appeared that LBVs can also
be the direct precoursors of supernova Type II explosions
\citep[e.g.,][and references therein]{Petrov2016}.

The nature of variable massive star winds is better understood for the
solar and subsolar metallicities, for which the wind power scales with metallicity Z,
indicating the dominant role of radiation pressure to ions of metals  \citep[e.g.][]{Vink2022}.
Extrapolation of such wind mechanism to very low metallicities assumes
no WR stars in the most metal poor starburst. However, the detection of WR
population in the most metal-poor BCGs  \citep[e.g.][]{Guseva2000}) evidence
that other mechanisms may operate for massive star winds at the very low metallicities.
Moreover, for the most dramatic events known as LBV 'giant eruptions', the
radiation pressure mechanism may be not suitable  \citep[e.g.][]{Smith2006}.

Hence, the study of individual very low-Z massive stars can be crucial for
the choice of their most reliable models. The understanding of processes in very
metal-poor LBVs is especially suitable since due to their highly
non-stationary  mass loss, their properties should be more sensitive to
the model assumptions.

The recent studies of massive stars with metallicities as low as Z = Z\sunn/10
in Sextans~A \citep{Lorenzo22, Schoote2022}, and Z = Z\sunn/20 in Leo~A \citep{Gull22}, represent
the important steps in understanding the properties of the low-metallicity massive stars.
However, the search for opportunities to study massive stars at even lower Z, say at Z\sunn/50 -- Z\sunn/40,
still remains highly challenging.

The number of individual very massive stars in galaxies with
the extremely low metallicity (Z $\lesssim$ Z\sunn/30,
corresponding to the host SF regions with the gas oxygen abundance of
12+log(O/H) $\lesssim$ 7.2 dex) is very limited in the local Universe.
Even more so it relates to such low-Z massive stars at the {\it advanced} evolutionary stages.
They, first of all, include a number of WR stars and red supergiants (RSGs) in the prototype XMP dwarf IZw18
\citep[e.g][]{IT97, Legrand97, Hirschauer24} at D $\sim$16~Mpc.

Another important object is the dwarf XMP galaxy DDO68 at D = 12.75~Mpc ($\mu$ = 30.53~mag)
\citep{Makarov17, Cannon14}, as derived via the TRGB method. In this galaxy, the lowest metallicity
LBV star was discovered (also known as DDO68-V1) \citep{LBV, IT09} along with many
identified individual supergiants \citep{DDO68LBV}.

\begin{figure*}
\centering
\includegraphics[angle=0,width=17.5cm, clip=]{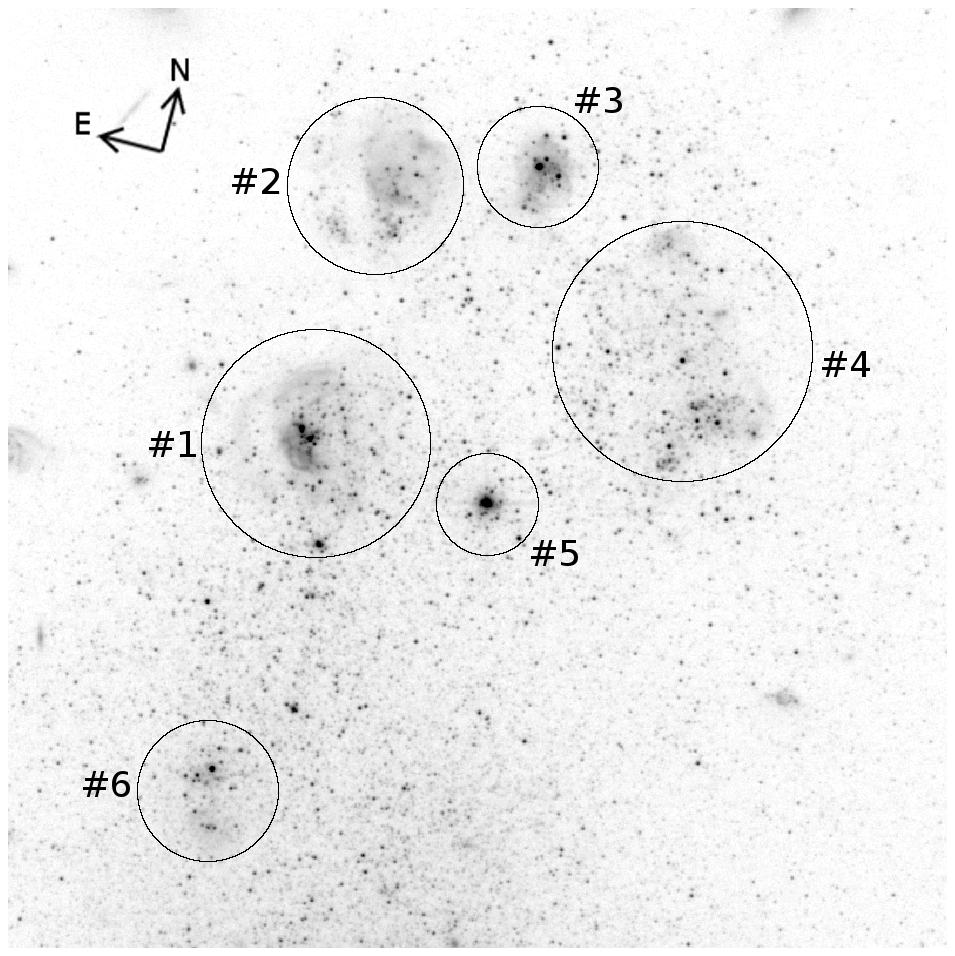}
\caption{
 The part of the HST image of DDO68 in F606W filter centred on the 'Northern Ring'
 (Knots 1--5), including  Knot~6 to the SW, with the used apertures superimposed.
 Image was acquired on 2010.05.01 for the HST program ID11578 (PI Aloisi).
 The  diameters of the apertures (in arcsec) are as follows: Knot~1 (10.0),  Knot~2 (8.0),  Knot~3 with DDO68-V1
 in the center (5.0),  Knot~4 (11.0),  Knot~5 (4.4),  Knot~6 (6.0).
}
	\label{fig:knot_1_6_V}
\end{figure*}

The Lynx-Cancer void galaxy DDO68 (UGC5340, VV542) is known as a peculiar
morphology, the almost record-low metallicity (average 12+log(O/H)=7.14) dIrr with
several prominent young star-forming (SF) regions. Most of these SF regions are
found at the periphery, mainly in the 'Northern ring' and the 'Southern tail'
\citep{DDO68, IT07}. These SF regions, named Knots
in \citet{DDO68}, look in fact as young OB associations hosting tens massive stars, with
the typical linear sizes of several tens pc. The only exception is the compact Knot~5,
which shows properties of a young globular cluster with the age of about 20~Myr.
See the HST image of Knots 1--6 in Fig.~\ref{fig:knot_1_6_V}.

\citet{Annibali2019} found the range of metallicities in DDO68, from 12+log(O/H) $\sim$7.3~dex near
the center of the main body, down to $\sim$7.0~dex in the southernmost SF region.
The importance of this XMP dwarf in the context of the future study of the lowest metallicity
massive stars was first emphasised about 20 years ago by \citet{IAUS232}.
Thanks to good luck, the unique Luminous Blue Variable star (LBV) was
discovered in the repeat observations of DDO68, in one of the most metal-poor SF regions in the local Universe
\citep{LBV,IT09}.

The issue of the extremely metal-poor massive star evolution and death remains
one of the clues for understanding of galaxy formation and evolution for
the time  of $\lesssim$1~Gyr since the Big Bang  \citep[e.g.,][]{Barkana01, Eldridge22}.
This is especially actual in the epoch of
active observational exploration of galaxies for the time of 0.5 -- 1~Gyr
since the Big Bang with the James Webb Space Telescope (JWST).
The great majority of these remote objects are star-forming galaxies.
They display a wide range of metallicities, including the lowest values, found in the
local Universe \citep[e.g.][]{Maseda23}.

While the state-of-art stellar evolution models, including those with the
fast rotation, have substantially advanced during the last decade
\citep[e.g.][and references therein]{Szecsi15, PARSEC, Sanyal17, Martins21}
the direct comparison  of the model predictions with properties
of real extremely metal-poor massive stars is still absent. The main reason
is the lack of such stars in the local Universe which would be
accessible for sufficiently detailed studies. One needs to wait for the
qualitative progress, expected for the upcoming extremely large next
generation optical telescopes. As a preliminary step for these future
studies, the search for such rare massive stars and examination of
their available properties, such as variability, appears valuable and
necessary.

As explained above, the main goal of the monitoring of the DDO68 'Northern Ring' region
is an attempt to examine the possible peculiarities in the light curve of the extremely
metal-poor LBV DDO68-V1. However, the images used for this task, cover the five other
SF regions (Knots), in which the HST data allow us to separate many other massive stars.
In particular, the list of the 50 brightest stars (supergiants with M$_{\rm V}$ $\leq$ --6.0~mag)
in these Knots was presented in \citet{DDO68LBV}.
Their sufficiently large-amplitude variability can, in principle, be detected
via variations of the integral light of their host SF regions.

The mentioned above SF regions, excluding Knot~5, are rather young, as follows from
their large values of the equivalent widths of the emission H$\beta$ \citep{DDO68}.
With the range of EW(H$\beta$) = 47 -- 250~\AA, the respective ages of the instantaneous
star-bursts correspond to the range of 3--7 Myr \citep{Starburst99}. So that
one can expect the population of massive OB stars at various evolutionary
stages still alive. Their non-stationary stellar wind activity can manifest as
a substantial light variability.

In this paper we present the most updated lightcurve for Knot~3, containing
the most metal-poor LBV star DDO68-V1. The data will be used in the forthcoming paper, examining
the LBV light variations after its so-called 'giant eruption' in 2010--2013 years.
More emphasis in the paper is given to the 'by-product' -- monitoring the light variations
in the rest Knots.
We present at the first time the magnitudes and lightcurves of Knots 1, 2, 4, 5 and 6 and
examine them for the possible signs of variability using the well-known $\chi$$^2$ criterion and
the so-called RoMS (Robust Median Statistics, \citet{Rose07, Burdanov14}).

The lay-out of the paper is as follows. In Sect.~\ref{sec:obs} we present
all the used observational data and briefly describe their reduction. In
Sect.~\ref{sec:results} the main results of data processing and analysis
are presented. Sect.~\ref{sec:vary} describes arrays of photometric serieses
for all Knots and their check for possible variability.
Sect.~\ref{sec:dis} is devoted to discussion of new results,
their comparison with the previous data and understanding them in a wider
context. In Sect.~\ref{sec:summ} we summarise the new results and draw
the main conclusions.
The adopted linear scale is 62~pc in 1~arcsec.

\section[]{OBSERVATIONAL DATA AND REDUCTION}
\label{sec:obs}

\subsection{BTA telescope data}
\label{ssec:BTA}

The imaging of the mentioned above SF regions (Knots 1--6) was conducted  with the multimode device
SCORPIO-1 \citep{SCORPIO} at the prime focus of the SAO 6-m telescope (BTA). The parameters of observations
are shown in the Journal (Tab.~\ref{Tab1_BTA}).  The details of observations are given in our papers PKP05;
\citet{LBV} and \citet{DDO68LBV}. SCORPIO-1 with the CCD detector 2K$\times$2K EEV~42-40 was used
in the  imaging mode. 
The images with the field of view with diameter of 6~arcmin, were obtained in the binning mode of 2$\times$2,
with the scale of  0.356~arcsec pixel$^{-1}$.

\subsection{SAO 1m telescope data}
\label{ssec:Zeiss}

We performed observations ($B$, $V$ and $R$ passbands) during the years 2016 - 2023  with
the 2048$\times$2048 CCD photometer on the 1-m telescope of SAO RAS.
The detector was CCD-EEV 42-40 with the full imaged field of
 $\sim$7.3~arcmin and the scale (after binning of 2$\times$2) of 0.432 arcsec/pixel.

The observations were conducted in the nights
with multi-program schedules. Therefore, the  images of DDO68 were obtained
mainly in $V$-band. When there was an opportunity, the images in $B$ and/or $R$-bands
were also acquired.
The log and parameters of observations are presented  in the general journal of observation
in Table~\ref{Tab1_BTA} and indicated in the 4-th column as '1m'.

\subsection{CMO 2.5m telescope data}
\label{ssec:CMO}

We performed observations (in V and R passbands) at 2.5-m telescope of Caucasian
Mountain Observatory (CMO) of Moscow State University in 2017--2019. We used
4296$\times$4102 NBI camera (made in the Niels Bohr Institute, Copenhagen) \citep{Shatsky2020}  mounted
in the Cassegrain focus of the telescope.
The camera has a 10$\times$10~arcmin field of view and scale of 0.16 arcsec/pixel.

Three observations of DDO68 were obtained during this period.
Two images were acquired in V-band. The first - in April 2017, with the total exposure time of 25 minutes,
and the second - in February 2018, with the total exposure time of 30 minutes.
The third image  was obtained in R band in January 2019 during the Moon time, with the total
exposure time of 30 minutes. See details in Table~\ref{Tab1_BTA}.

\begin{table}
\begin{center}
\caption{Journal of SAO and CMO observations of DDO68}
\label{Tab1_BTA}
\hoffset=-2cm
\begin{tabular}{l|l|r|l|c|} \hline  \hline \\ [-0.2cm]
\MC{1}{c|}{Date} &
\MC{1}{c|}{Band}&
\MC{1}{c|}{Expos.}&
\MC{1}{c|}{Tel} &
\MC{1}{c|}{$\beta$\arcsec} \\

\MC{1}{c|}{ } &
\MC{1}{c|}{ } &
\MC{1}{c|}{time, s}&
\MC{1}{c|}{ } &
\MC{1}{c|}{}  \\

\MC{1}{c|}{(1)} &
\MC{1}{c|}{(2)} &
\MC{1}{c|}{(3)} &
\MC{1}{c|}{(4)} &
\MC{1}{c|}{(5)} \\
\\[-0.2cm] \hline \\[-0.2cm]
 2005.01.12  & $V,R$    & 1800,1800    & BTA & 1.7 \\ 
 2009.01.21  & $V$      & 300          & BTA & 1.3 \\ 
 2015.01.14  & $B,V,R$  & 300,360,360  & BTA & 2.1 \\ 
\hline 
 2016.03.07  & $V$      &         4200 & 1m  & 2.1 \\ 
 2016.03.08  & $B,V,R$  &1200,1200,1200& 1m  & 2.1 \\ 
 2016.04.07  & $V$      &         4200 & 1m  & 2.2 \\ 
 2016.05.17  & $V$      &         4200 & 1m  & 2.2 \\ 
 2016.10.22  & $V$      &         3600 & 1m  & 1.8 \\ 
 2016.11.24  & $B,V,R$  &3600,3600,3600& 1m  & 1.9 \\ 
 2016.12.24  & $V$      &         3600 & 1m  & 2.6 \\ 
 2016.12.31  & $V$      &         5400 & 1m  & 1.6 \\ 
 2017.04.18  & $V$      &         1500 & CMO & 1.7 \\ 
 2017.05.29  & $V$      &         1200 & 1m  & 1.7 \\ 
 2017.11.16  & $B,V,R$  &  600,600,600 & BTA & 1.9 \\ 
 2018.02.19  & $V$      &         1800 & CMO & 1.8 \\ 
 2018.04.05  & $B,V,R$  &2100,2400,3900& 1m  & 1.8 \\ 
 2018.04.30  & $V,R$    &2400,2400     & 1m  & 1.7 \\ 
 2018.10.11  & $V$      &3300          & 1m  & 1.6 \\ 
 2019.01.18  & $R$      &1800          & CMO & 1.0 \\ 
 2019.02.03  & $V$      &3600          & 1m  & 1.6 \\ 
 2019.10.26  & $B,V,R$  &600,600,600   & BTA & 1.2 \\ 
 2019.11.25  & $V$      &3600          & 1m  & 1.6 \\ 
 2020.01.19  & $R$      &1200          & BTA & 1.2 \\ 
 2020.01.20  & $B,V$    &600,600       & BTA & 1.5 \\ 
 2020.03.04  & $V,R$    &2400,1500     & 1m  & 1.7 \\ 
 2020.04.26  & $R$      &600           & BTA & 2.0 \\ 
 2020.11.11  & $V,R$    &600,900       & BTA & 1.5 \\ 
 2021.05.14  & $V$      &2400          & 1m  & 2.0 \\ 
 2021.05.16  & $R$      &1200          & 1m  & 2.0 \\ 
 2021.12.02  & $V,R$    &720,720       & BTA & 1.6 \\ 
 2022.04.26  & $V$      &3000          & 1m  & 1.7 \\ 
 2022.10.25  & $V,R$    &2700,1500     & 1m  & 2.5 \\ 
 2022.12.20  & $B,V,R$  &720,720,720   & BTA & 1.8 \\ 
 2022.12.25  & $B,V,R$  &1500,3600,2700& 1m  & 2.2 \\ 
 2023.01.23  & $B,V,R$  &3600,3600,3600& 1m  & 2.1 \\ 
 2023.03.18  & $V$      &2700          & 1m  & 1.9 \\ 
 2023.10.12  & $V$      &3600          & 1m  & 1.7 \\ 
 2023.10.22  & $V$      &900,900       & BTA & 1.9 \\ 
\hline \hline \\[-0.2cm]
\end{tabular}
\end{center}
\end{table}

\subsection{Literature and archival data}

\begin{table}
\begin{center}
\caption{Summary of archive data for DDO68}
\label{Tab2_external}
\hoffset=-2cm
\begin{tabular}{l|l|l|r|c} \hline  \hline \\ [-0.2cm]
\MC{1}{c|}{Date} &
\MC{1}{c|}{Telescope} &
\MC{1}{c|}{Band}&
\MC{1}{c|}{Expos.}&
\MC{1}{c|}{$\beta$\arcsec} \\

\MC{1}{c|}{ } &
\MC{1}{c|}{ } &
\MC{1}{c|}{ }&
\MC{1}{c|}{time, s} &
\MC{1}{c|}{} \\

\MC{1}{c|}{(1)} &
\MC{1}{c|}{(2)} &
\MC{1}{c|}{(3)} &
\MC{1}{c|}{(4)} &
\MC{1}{c|}{(5)} \\
\\[-0.2cm] \hline \\[-0.2cm]
 1988.02.14  & 3.5m CA   & $B,R$    & 900,600       & 1.1  \\ 
 1994.05.02  & 2.54m INT & $R$      & 1200          & 1.9  \\ 
 1995.02.07  & 2.56m NOT & $B,V$    & 300,300       & 0.9  \\ 
 1997.12.23  & 10m KeckII& $V$      & 600           & 1.0  \\ 
 1998.02.17  & 4m KPNO   & $B,V$    & 600,600       & 2.5  \\ 
 2000.04.07  & 1.8m VATT & $B,V,R$  & 600,480,360   & 1.7  \\ 
 2004.04.16  & 2.5m APO  & $g,r,i$  & 57,57,57      & 1.4  \\ 
 2007.02.09  & 2.1m KPNO & $g,i$    & 1800,1800     & 0.9  \\ 
 2010.05.02  & 2.6m HST  & $V,I$    & 7644,7644     & 0.1  \\ 
 2011.03.05  & 0.9m KPNO & $B,V,R$  & 900,600,420   & 1.4  \\ 
 2013.02.17  & 0.9m KPNO & $B,V,R$  & 1200,1200,1200& 1.9  \\ 
\hline \hline \\[-0.2cm]
\end{tabular}
\end{center}
\end{table}

In addition to our new photometry obtained with SAO 6m and 1m telescopes and 2.5m telescope of
MSU, presented in Table~\ref{Tab1_BTA}, we use also the available archive CCD images, which we analysed
in the similar way.
In Table~\ref{Tab2_external} we present the summary of imaging observations of DDO68
based on archives of nine ground-based telescopes world-wide and the HST data. Below we briefly
comment on the origin of this data and refer to the papers, in which the respective programs
and  related issues are described in more detail.

\subsubsection{Calar Alto 3.5-m telescope data}

The observations of galaxy DDO68  were collected with the Prime focus CCD camera of
the Calar Alto 3.5-m telescope in dark time on the night 1988.02.14 
under photometric conditions by \citet{HS95}.
A RCS high resolution chip was used with the field of view of 256 by
160~arcsec and a resolution of 0.5 arcsec/pixel.
For DDO68, the images were obtained in B-band and R-band, with the exposure
times of 900~s and 600~s, respectively. 
Seeing was 1.1~arcsec.
The program observations were accompanied by the appropriate
CCD calibration observations (bias and dark frames and sky flats).
After de-biasing, the dark current was subtracted, bad columns were
removed, and the frames were flat-field corrected.

\subsubsection{Isaac Newton 2.54-m telescope data}

The image of DDO68 was obtained on 1994.02.05 by \citet{SB02}
with Johnson-Cousins R-band filter at seeing of 1.86~arcsec  
with the exposure time of 1200~s.
The detector was CCD-EEV 4K$\times$2K (EEV XHIP) with scale of 0.549 arcsec/pixel.

\subsubsection{Nordic Optical (NOT) 2.56-m telescope data}

Observations of DDO68 were carried out on 1995.02.07 on the program of
\citet{Makarova98} with the NOT 2.56-m telescope (La Palma). Images in
B and V bands were obtained with the exposure times of 300~s and 300~s, and seeings
of  0.9 and 0.7~arcsec, respectively.
A CCD camera with a TEC 1k$\times$1k chip provided a field of view of
3$\times$3 arcmin, with the image scale of 0.176 arcsec/pixel.

\subsubsection{Keck~II  10-m telescope data}

Observations of DDO68 were carried out  on the night of 1997.12.23 with the Low
Resolution Imaging Spectrometer (LRIS) at the f/15
Cassegrain focus of the Keck II telescope by \citet{Mendez02}.
The used back-side-illuminated Tektronix 2048$\times$2048 CCD detector 
had an image scale of 0.215 arcsec/pixel and a field of view of
6$\times$8 arcmin. The images were obtained in V-band with the exposure time
of 600~s, seeing of 1.0 arcsec and with  the airmass of 1.02.

\subsubsection{KPNO  4-m telescope data}

The KPNO 4-m telescope images for DDO68 were obtained on the night 1998.02.17 in the frame
of the program from the paper by \citet{HE06}.
The 2K$\times$2K Tektronix CCD images in B and V bands were acquired at seeing
of 2.5~arcsec with the exposure time of 600~s. Scale at detector was 0.26 arcsec/pixel.

\subsubsection{VATT  1.8-m telescope data}

Images of DDO68 on 2000.04.07    
were obtained with the Vatican 1.8~m telescope (VATT, see for more detail in
 \citet{Taylor05}), located at Mt. Graham International Observatory (MGIO,
Arizona, USA), with the 2K$\times$2K  Direct Imager.
Exposure times in B, V, R filters were 600, 480, 360~s, respectively,
at seeing of 1.5 -- 1.8~arcsec. Scale on the detector was 0.374 arcsec/pixel.

\subsubsection{KPNO 2.1-m telescope data}

The images of DDO68 on 2007.02.09  were obtained with the KPNO 2.1-m telescope
\citep{IT09} in the SDSS filters g and i, with the exposure time for both
filters of 3$\times$600~s.
Airmass during observations was 1.05, with seeing of 0.9~arcsec.
The used detector was Tektronix 2K$\times$2K imager with the scale of 0.27 arcsec/pixel.
The obtained magnitudes in g and i filters were transformed to that in V-band with
the equations of \citet{Lupton05}.

\subsubsection{The 2.6-m Hubble Telescope data}

Imaging with the HST were obtained on 2010.04.30 (F814W) and 2010.05.02 (F606W)
within the Program ID GO 11578 (PI A.Aloisi). The Advanced Camera for Surveys was used.
The data processing and photometry for Knot~3 were already described in \citet{DDO68LBV}.
For the other Knots, all aperture photometry was done similarly, with the same diameters of
aperture as for the ground-based observations. The relatively large error of V-band magnitude
of 0.08~mag is due to the related uncertainty of the transform from the HST filter magnitudes
to Johnson - Cousins V-band magnitudes. For the concrete HST image, from which we derive V-band
magnitudes of all our Knots, this rms transform 'error' will cause the same systematic shift
for all Knots. If the value of such a shift is comparable to the amplitude of the real variations
of $\sim$0.1~mag, this can manifest in the lightcurves of the Knots as a correlated uprise or
decline.

\subsubsection{KPNO 0.9-m telescope data}

Observations of DDO68 with the Kitt Peak National Observatory's
0.9 meter telescope were carried out on the nights 2011.03.05 and 2013.02.17
in B, V, R bands \citep{Cook14}.
No publication was found in the literature
related to the latter data.      
The S2KB detector was a 2048$\times$2048 CCD mounted at the f/7.5 focus of the telescope.
The field size was approximately of 20 arcmin on a side, with the scale of 0.6 arcsec/pixel.

\subsection{Details of photometry}
\label{ssec:photo_detail}

To derive the total magnitudes of the DDO68 Knots 1--6 in the 'Northern ring',
we used MIDAS\footnote{MIDAS is an acronym for the European Southern Observatory package -- Munich
Image Data Analysis System}-based  programs to perform the standard photometric pipeline with bias
subtraction, flat-fielding and illumination correction.
The background was approximated by the 2D polynomial of the 2nd degree. This was fitted on the data
for $\sim$80 small apertures around the Northern Ring, with the total area close to that of the Ring .

 We neglect in the further estimates of the magnitude errors the component from the dark current and
the read-out noise, since they are many times smaller than the contribution of Poisson noise from
the sky background and an object.
Adopting that the standard deviation of the Poisson process with N counts\footnote{Here N is
the detector electron number counts, that is ADU counts multiplied by 'gain'.}  is $\sqrt{N}$,
the relative error of the object (Knot) flux is read as:
   $\sqrt{N_{\rm sky+obj}}$ / $N_{\rm obj}$
where $N_{\rm sky}$ - number counts of sky within the used aperture,
$N_{\rm sky+obj}$ - number counts of sky together with the object.
$N_{\rm obj}$ = $N_{\rm sky+obj}$ -- $N_{\rm sky}$, the difference number counts, the pure object counts.
The subtraction of $N_{\rm sky}$, fitted by the above polynomial, increases the Poisson standard deviation
of the signal within the aperture on the object only by a few per cent.

For instrumental magnitudes, \mbox{$m_{\rm inst}$ = --2.5~log($N_{\rm obj}$)},
their respective errors $\sigma_{\rm inst}$,
assuming that they are small, are calculated as follows:
 \mbox{$ \sigma_{\rm inst}$ = 2.5~log~(1 + $\sqrt{N_{\rm sky} + N_{\rm obj}}$ / $N_{\rm obj})$}~~~~(1).

The photometry of DDO68 Knots 1--6 was based on the local standard stars. To transform
the instrumental magnitudes to the apparent magnitudes in the respective bands,
we used the zero-points determined for every image through the measurements
of 5--6 or more local standard stars. These sufficiently bright stars
($g$ = 17.9 -- 21.1~mag) were selected in the Sloan Digital Sky Survey
 \citep[SDSS DR7;][]{DR7} images in the close vicinity of DDO68 so that they fall within  the
typical fields of view  of our used images ($\gtrsim$6~arcmin across).
Their $g,r,i$ magnitudes were measured independently with the aperture photometry and
 were transformed to the Johnson-Cousins $B,V,R$
magnitudes according to the \citet{Lupton05} relations. These local standard stars are presented
in Table~\ref{tab:local_standards}, where we give their SDSS names, $g,r,i$ magnitudes
and the adopted Johnson-Cousins $B,V,R$ magnitudes.

The apertures, used for the local standards, had radii from 2.1 to 3.1~arcsec and were adjusted in
order to minimise the contribution of the local faint neighbours.
The related small corrections to the asymptotic value of magnitude, depending on the seeing, were applied similar
to those for the Knots. To find these corrections, we used the original HST images and smoothed them with
the Gaussian filter corresponding to seeings between 1.0 to 2.5 arcsec. 
In each case, to obtain the total magnitude of the studied Knots, we used the round apertures
 with the fixed diameters, indicated in Table~\ref{tab:sum_knots}.

\subsection{Errors budget}
\label{ssec:errors}

In Sect.~\ref{sec:vary} we perform the statistical checks of the Null hypothesis about the absence of variability in
the photometric time serieses of the studied DDO68 Knots.
These tests are based on the measurement errors, assuming that we have their correct estimates and
all systematic and non-random factors are excluded. We briefly summarise the error estimate procedure and outline
the issues which could potentially affect the errors of photometry.

The main advantage of our photometry was the use of the  number of local standards what allowed us to define
the zero-points (as the weighted mean on several standards) for every measurement.
The errors of these weighted means in V-band fall between 0.002 and 0.016~mag, with the median of $\sim$0.007~mag,
and 3/4 of them of smaller than 0.012~mag. This estimate of the error of a zero-point was combined
in quadrature with the relative error of the measured flux for the respective Knots.
The errors for the zero-point in B-band fall between 0.002 and 0.023~mag, with the median error of $\sim$0.008~mag.
Similarly, for R-band, the errors of the zero-point fall between  0.002 and 0.018~mag, with the median of $\sim$0.008~mag.

As mentioned in Sect.~\ref{ssec:photo_detail}, the relative error of the measured flux is determined
by the Poisson noise term shown under $\sqrt{N}$ in formula (1) above.
For all Knots, the term $N_{\rm sky}$ is larger than $N_{\rm obj}$ by a factor of $\sim$3--10 in B-band,
$\sim$5--10 in V-band and $\sim$15 in R-band. Therefore the contribution from an object to the Poisson noise
is typically smaller than 10--20 per cent.
Subtraction of the sky background, approximated by the 2D polynomial,
as described above, increased the standard deviation in the resulting image only by a few per cent. So, we ignored
this small addition of the noise.

For the typical time exposures per filter of 10~minutes at BTA, for case of Knot~3, the faintest of all six Knots
(see Table~\ref{tab:sum_knots}), the respective numbers of
$N_{\rm sky}$ are $\sim$3.6$\times 10^{5}$, $\sim$5.6$\times 10^{5}$ and $\sim$1.1$\times 10^{7}$
in B, V and R-bands, respectively. The respective counts of $N_{\rm obj}$ in B, V and R bands
were $\sim$1.3$\times 10^{5}$, $\sim$1.0$\times 10^{5}$ and $\sim$6.6$\times 10^{5}$.
The typical error of the instrumental magnitudes,
of $\sigma_{\rm inst}$ are 0.006, 0.009 and 0.006~mag in the respective bands.

For observations at the SAO 1m telescope, with the typical exposure time of 60~minutes, the respective numbers
of $N_{\rm sky}$ are $\sim$4.8$\times 10^{5}$, $\sim$7.5$\times 10^{5}$ and $\sim$1.8$\times 10^{6}$ in B, V
and R-bands, respectively.
The respective counts of $N_{\rm obj}$ in B, V and R band were $\sim$5.5$\times 10^{4}$, 7.0$\times 10^{4}$
and 6.5$\times 10^{4}$.
The related typical errors are $\sigma_{\rm inst}$ are 0.014, 0.013 and 0.024~mag in the respective bands.

The zero-points were adopted as the weighted means on all available local
standards on the used image. Their errors were accepted as the errors of
these weighted means. For the SAO 1-m telescope, these errors were in the
range of 0.010 to 0.018 mag, depending on the band and the night, while for images obtained
with the 6m telescope, the uncertainties of zero-points were a factor of 2 smaller.
The errors of measured apparent magnitudes for Knots 1--6  were calculated as
the sums in quadrature of the Poisson noise standard deviation and the error
of zero-points.

We identified the only factor leading to the systematic underestimate of the Knots' fluxes.
Depending on the value of seeing $\beta$, defined as FWHM of a star profile, a small fraction of
a Knot total flux appears lost for photometry with the fixed size aperture.
To evaluate this fraction and perform the respective correction, we proceed as follows.
We used the HST images obtained in December 2017 (GO program 14716, PI: F.~Annibali). We performed the
Gaussian filter smoothing by the MIDAS procedure 'filter/gauss' on the HST F606W image
to produce images with the seeings $\beta$, corresponding to those we had on the images obtained
with the ground-based telescopes.

We compare the results of aperture photometry for non-smoothed and smoothed images
with the fixed sizes of aperture. From this we evaluate corrections for the measured
magnitudes of each Knot with the respective size of aperture, depending on the actual seeing. These
corrections are larger for the smaller apertures. They vary from 0.01--0.02~mag for $\beta \sim$1.0--1.2~arcsec,
to 0.04~mag for $\beta \sim$1.9~arcsec, to 0.06~mag for $\beta$ of $\sim$2.2~arcsec, and up to
0.07~mag for the few occasional images with   $\beta$ of $\sim$2.5~arcsec. These corrections were applied to
each magnitude estimate in Tables~\ref{tab:Knot1}--\ref{tab:Knot6}.

\begin{table*}
\begin{center}
\caption{Summary of properties of local standards}
\label{tab:local_standards}
\begin{tabular}{l|l|r|l} \hline  \hline \\ [-0.2cm]
\MC{1}{c|}{SDSS Name} &
\MC{1}{c|}{SDSS~$g,r,i$}&
\MC{1}{c|}{$B,V,R$}&
\MC{1}{c|}{$\sigma_{\rm g}$, $\sigma_{\rm r}$, $\sigma_{\rm i}$} \\

\MC{1}{c|}{ } &
\MC{1}{c|}{mag}&
\MC{1}{c|}{mag} &
\MC{1}{c|}{mag} \\

\MC{1}{c|}{(1)} &
\MC{1}{c|}{(2)} &
\MC{1}{c|}{(3)} &
\MC{1}{c|}{(4)} \\
\\[-0.2cm] \hline \\[-0.2cm]
%
 J095633.91+284953.9 & 21.08, 20.45, 20.01 & 21.50, 20.71, 20.23 & 0.03, 0.03, 0.04   \\ %
 J095635.72+285042.6 & 20.54, 20.18, 20.16 & 20.87, 20.32, 20.02 & 0.03, 0.03, 0.03   \\ %
 J095639.60+285202.4 & 19.66, 18.21, 17.47 & 20.34, 18.82, 17.85 & 0.01, 0.01, 0.01   \\ %
 J095639.81+285207.7 & 19.09, 17.93, 17.51 & 19.67, 18.42, 17.62 & 0.01, 0.01, 0.01   \\ %
 J095643.53+284747.6 & 18.66, 18.09, 17.84 & 19.06, 18.33, 17.87 & 0.01, 0.01, 0.01   \\ %
 J095646.63+284928.3 & 17.91, 17.22, 16.98 & 18.36, 17.51, 17.00 & 0.01, 0.01, 0.01   \\ %
 J095648.62+284854.3 & 20.81, 19.70, 19.17 & 21.38, 20.16, 19.40 & 0.02, 0.02, 0.02   \\ %
 J095649.01+284911.2 & 20.35, 19.03, 18.24 & 20.98, 19.58, 18.68 & 0.02, 0.01, 0.01   \\ %
 J095649.51+285236.6 & 20.59, 20.34, 20.04 & 20.90, 20.44, 20.20 & 0.03, 0.03, 0.04   \\ %
 J095650.57+284950.1 & 20.66, 19.20, 18.35 & 21.34, 19.81, 18.83 & 0.02, 0.01, 0.01   \\ %
 J095652.10+285041.6 & 21.77, 20.64, 19.07 & 22.38, 21.12, 20.32 & 0.05, 0.04, 0.02   \\ %
 J095656.49+284911.7 & 18.99, 18.64, 18.56 & 19.33, 18.78, 18.47 & 0.01, 0.01, 0.01   \\ %
\hline \hline \\[-0.2cm]
\end{tabular}
\end{center}
\end{table*}

\begin{figure*}
  \centering
 \includegraphics[angle=-90,width=11.0cm,clip=]{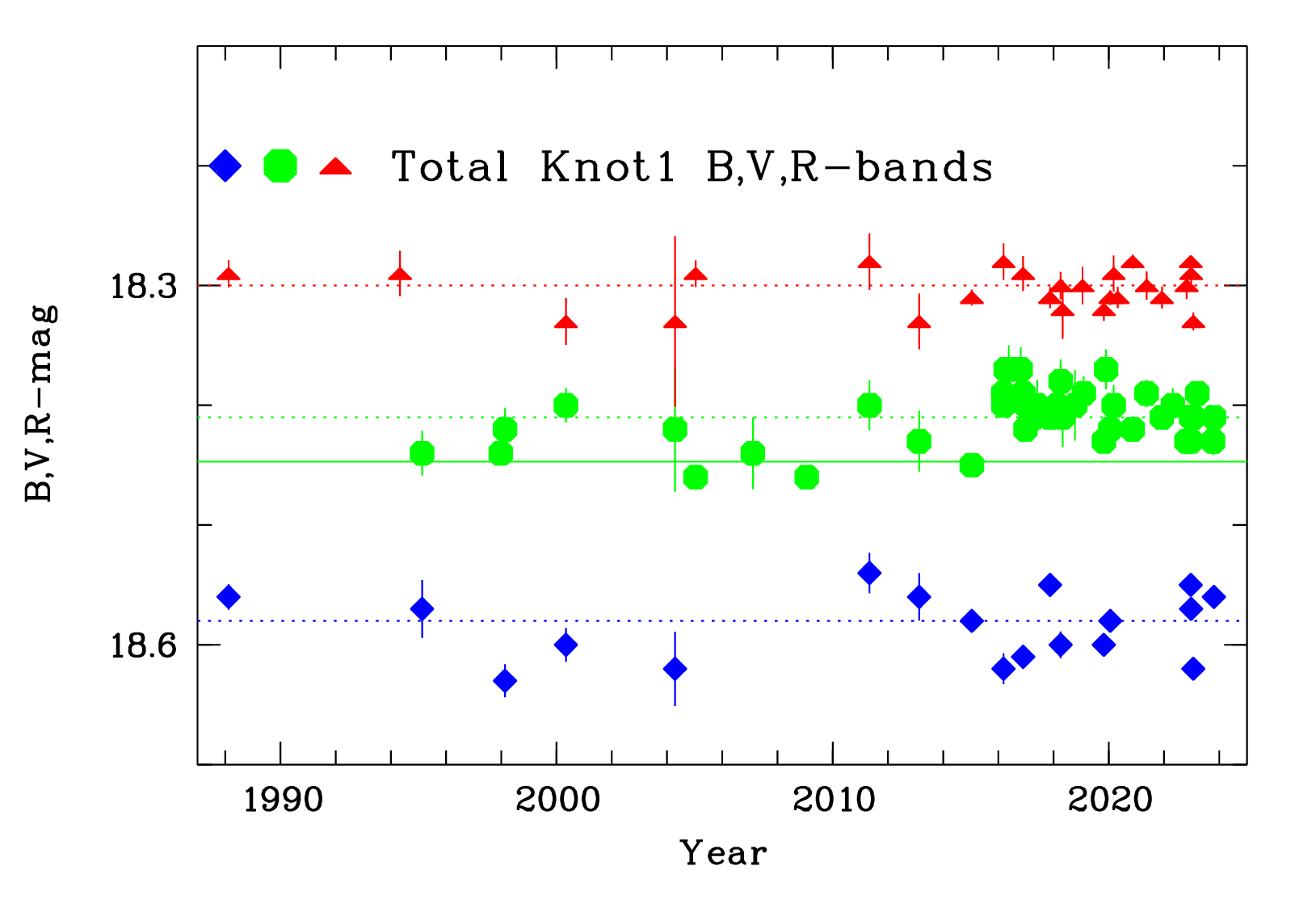}
 \includegraphics[angle=-90,width=11.0cm,clip=]{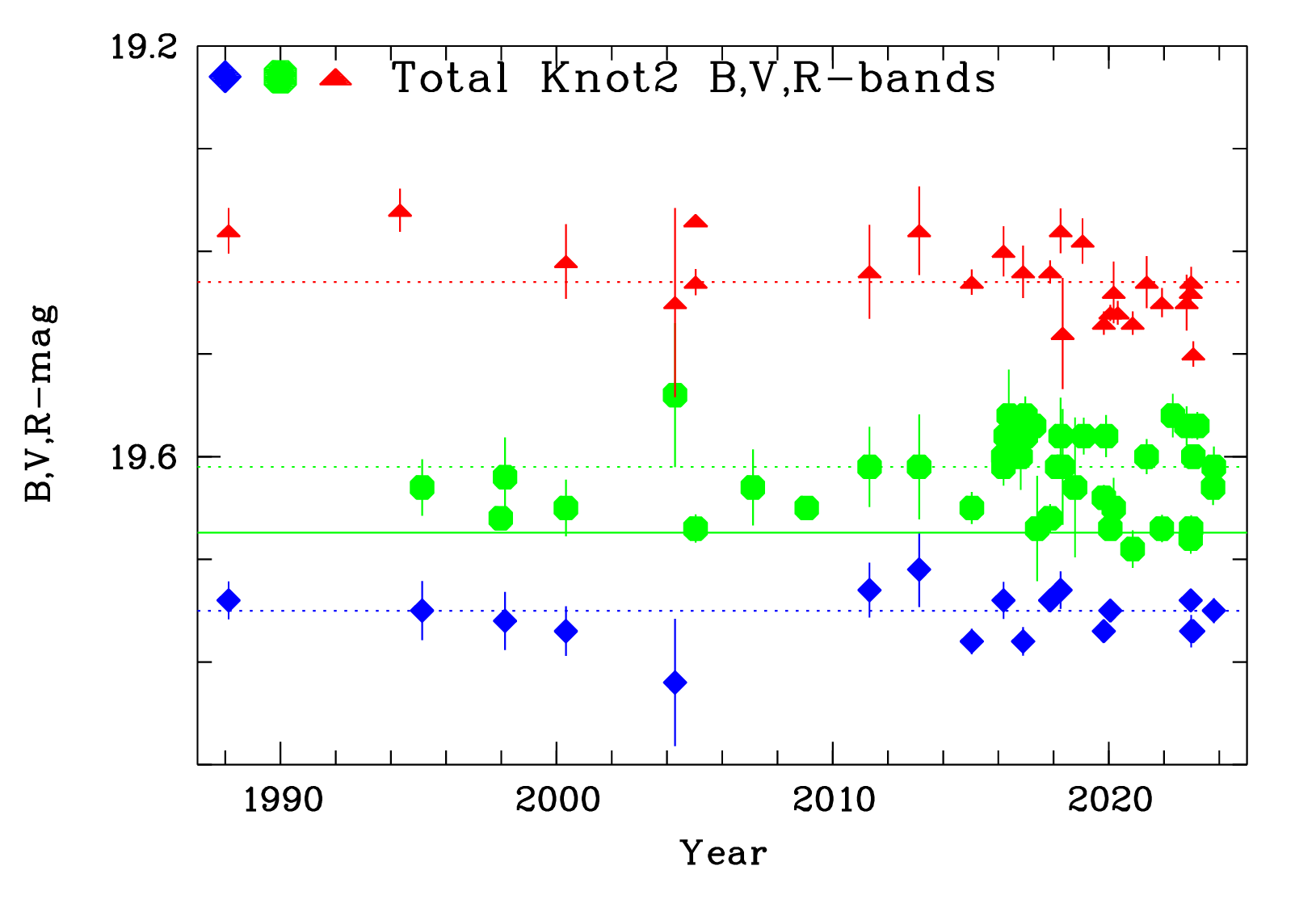}
 \includegraphics[angle=-90,width=11.0cm,clip=]{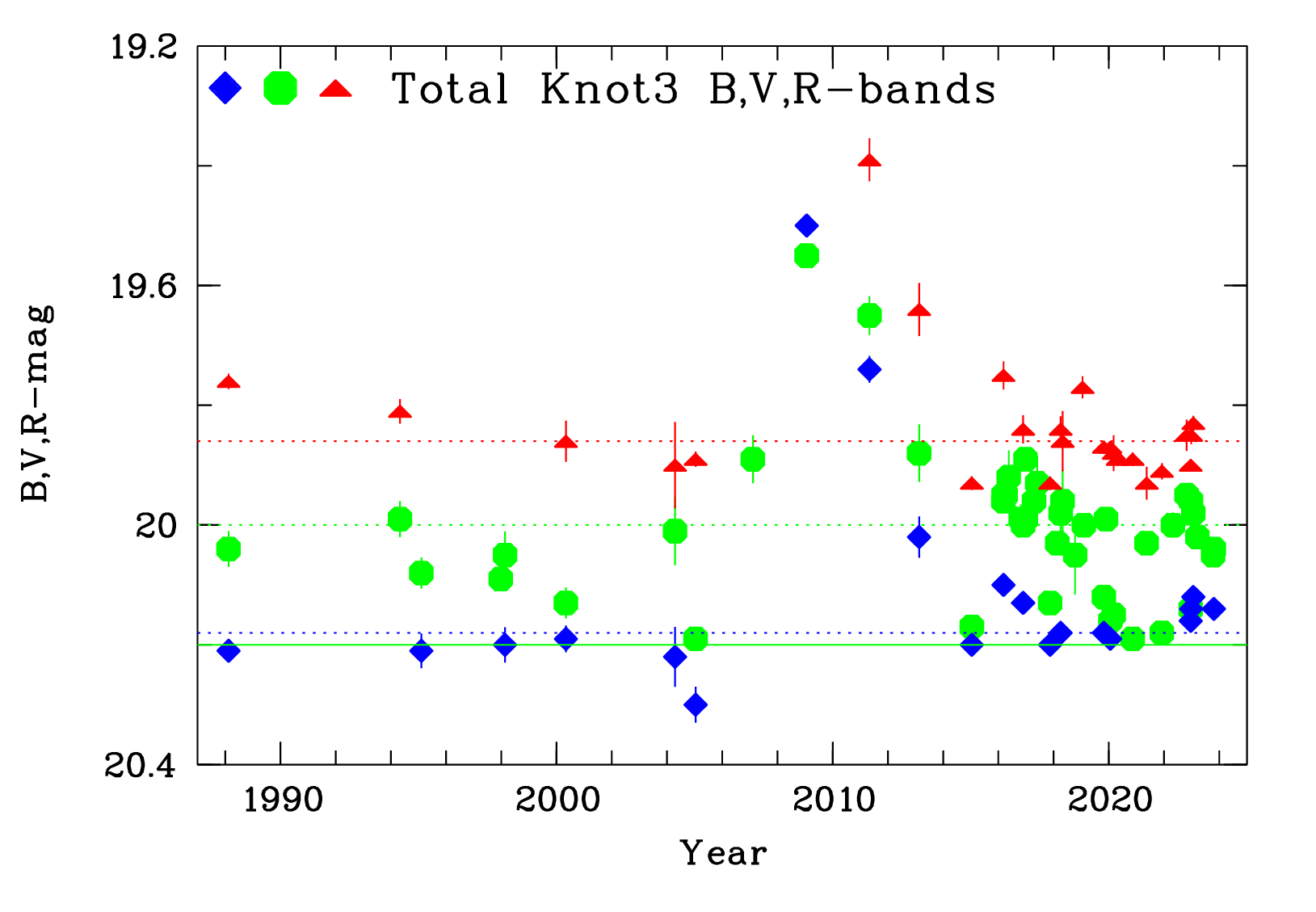}
  \caption{
From top to bottom: Lightcurves in $B,V,R$-bands of Knots 1, 2, 3.
Dashed lines show the median magnitudes to help the eye.
}
  \label{fig:lcurves_K1-K3}
 \end{figure*}

\section[]{RESULTS}
\label{sec:results}

\begin{figure*}
  \centering
 \includegraphics[angle=-90,width=11.0cm,clip=]{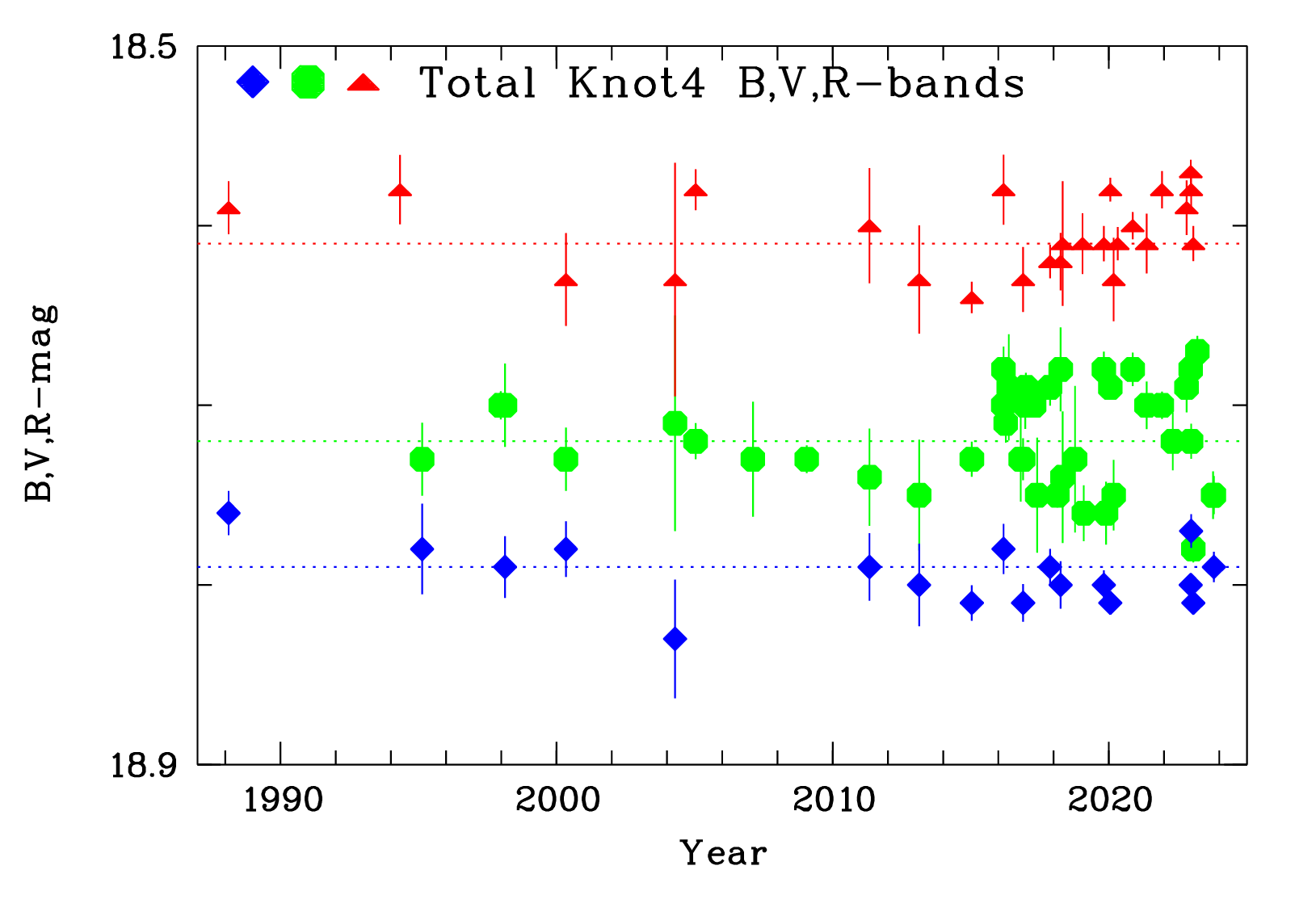}
 \includegraphics[angle=-90,width=11.0cm,clip=]{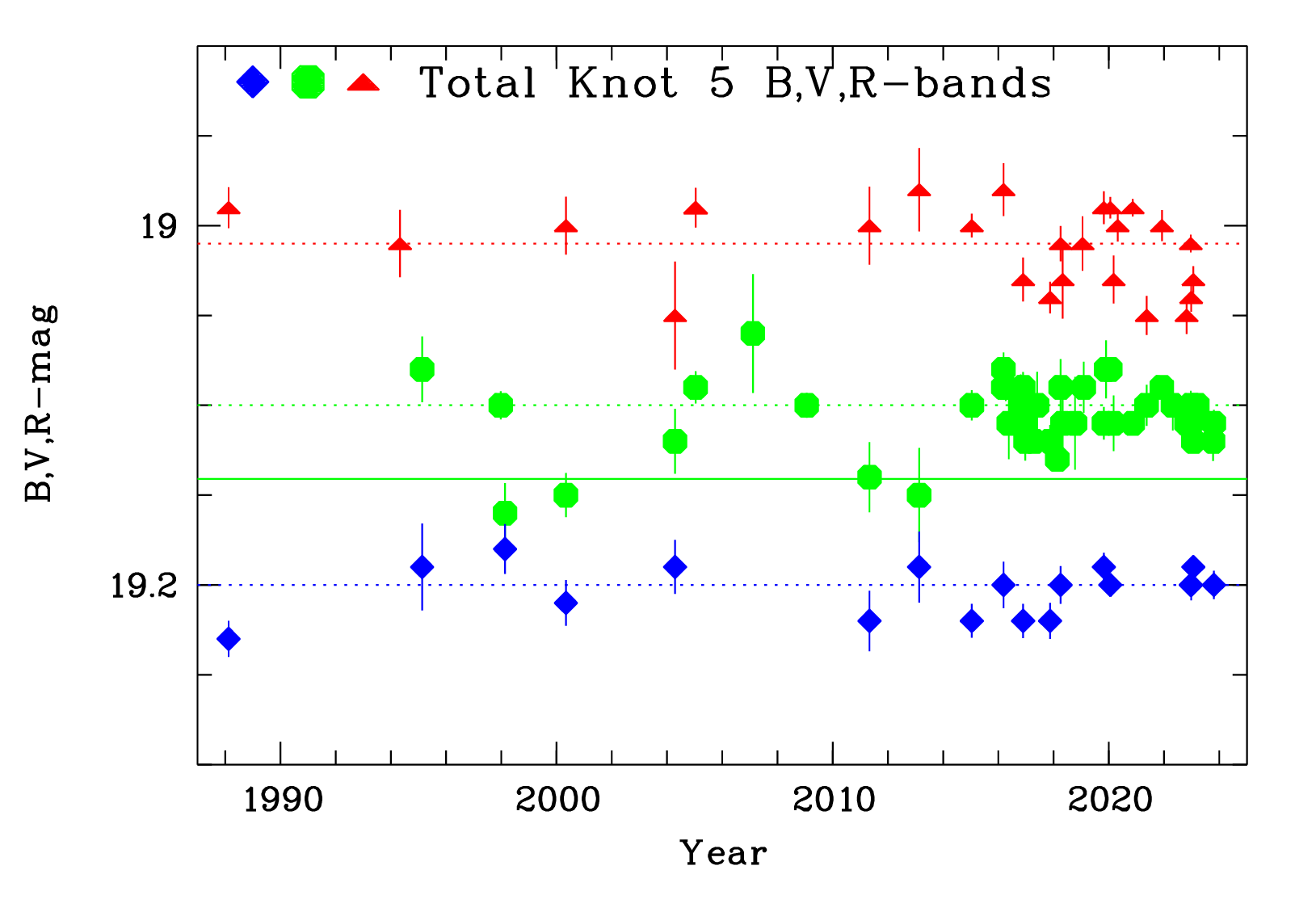}
 \includegraphics[angle=-90,width=11.0cm,clip=]{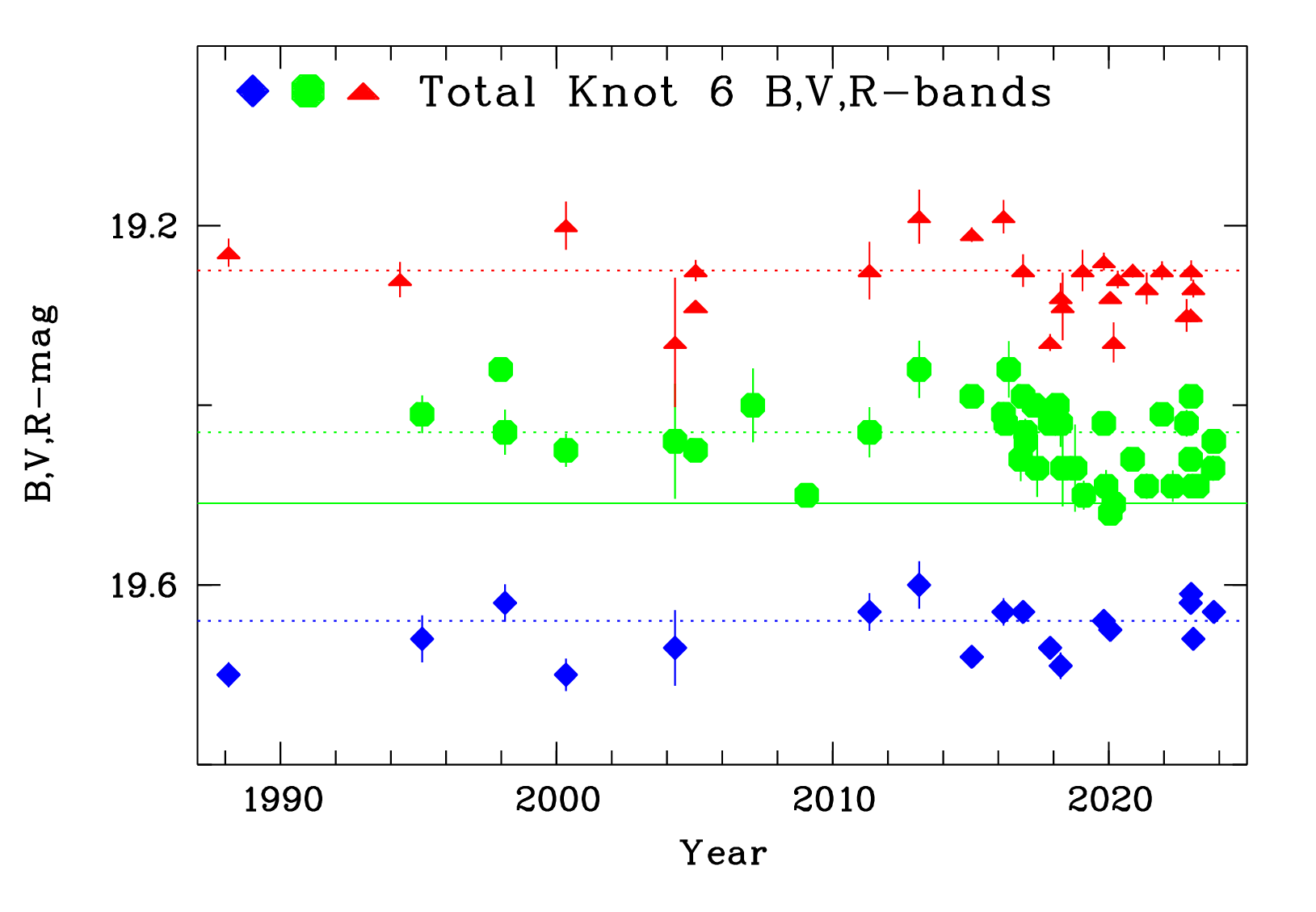}
  \caption{
Lightcurves of Knots~4, 5, 6 similar to those in Fig.~\ref{fig:lcurves_K1-K3}.
Dashed lines show the median magnitudes for each band. Solid lines for $V$ 
show the adopted minimal levels. They are defined to be consistent
with the maximal possible number of points at this brightness level.
}
  \label{fig:lcurves_K4-K6}
 \end{figure*}
 
In Table~\ref{tab:sum_knots} we briefly summarise results of photometry of all six Knots
for the most numerous data, collected for V-band, including their median magnitudes and the estimated amplitudes
 of variability in this band. The more advanced analysis of the potential variability and the related parameters
are presented in Sect.~\ref{sec:vary}.

The resulting lightcurves for all 6 knots in B, V, R bands are combined
in Figs.~\ref{fig:lcurves_K1-K3} and \ref{fig:lcurves_K4-K6} and discussed in Sect.~\ref{sec:dis}.
In Figures~\ref{fig:lcurves_K1_var} to \ref{fig:lcurves_K6_var}, we show the lightcurves of the residual light
for each Knot in V-band, that is the observed light minus that of the minimal level. We show these residual lightcurves
in three time intervals in order one could better see variations at various time-scales.

In Tables~\ref{tab:Knot1}--\ref{tab:Knot6} (Appendix~A, Sect.~\ref{sec:supp}), we present the data on
the individual measurements in each band of B, V and R,
with the corrections related to the effect of seeing on the loss of light entering the aperture. This issue
is described in detail in Sect.~\ref{ssec:errors}.

\begin{table}
\begin{center}
\caption{Summary of properties of Knots 1--6}
\label{tab:sum_knots}
\begin{tabular}{l|l|r|l|r} \hline  \hline \\ [-0.2cm]
\MC{1}{c|}{Name} &
\MC{1}{c|}{Coord. of Centre} &
\MC{1}{c|}{Aperture}&
\MC{1}{c|}{Median}&
\MC{1}{c|}{$\delta V$} \\

\MC{1}{c|}{ } &
\MC{1}{c|}{J2000 } &
\MC{1}{c|}{diameter}&
\MC{1}{c|}{$V$-mag} &
\MC{1}{c|}{mag} \\

\MC{1}{c|}{(1)} &
\MC{1}{c|}{(2)} &
\MC{1}{c|}{(3)} &
\MC{1}{c|}{(4)} &
\MC{1}{c|}{(5)} \\
\\[-0.2cm] \hline \\[-0.2cm]
 Knot~1   & J095646.73+285008.9 & 10.0$"$ & 18.41  & $\sim$0.09 \\ %
 Knot~2   & J095646.69+285020.8 &  8.0$"$ & 19.61  & $\sim$0.13 \\ %
 Knot~3$^{*}$ & J095646.25+285023.5 &  5.0$"$ & 20.00  & $\sim$0.30 \\ %
 Knot~4   & J095645.60+285015.9 & 11.0$"$ & 18.72  & $\sim$0.11 \\ %
 Knot~5   & J095646.15+285009.9 &  4.4$"$ & 19.10  & $\sim$0.08 \\ %
 Knot~6   & J095646.78+284954.8 &  6.0$"$ & 19.44  & $\sim$0.16 \\ %
\hline \hline \\[-0.2cm]
\multicolumn{5}{l}{$^{*}$ $\delta V$ during the period after the LBV 'giant eruption' event. } \\
\end{tabular}
\end{center}
\end{table}

\section[]{EXAMINATION OF POSSIBLE VARIABILITY}
\label{sec:vary}

For Knot~3, outside the time interval of the 'giant eruption' event, discussed in \citet{DDO68LBV}
and \citet{LBV_giant}, the variability with the amplitude of $\sim$0.3~mag
is quite evident and does not require special tests. As for the other Knots,
the light variations are significantly less pronounced.
Therefore, we need to use the known tests to check, whether the
observed scatter in the measured magnitudes indicates the real variations,
or they are consistent with the scatter due to the adopted observational errors.

As one can see from the data in Tables~\ref{tab:Knot1}--\ref{tab:Knot6}, due to the various
reasons, the photometry data in V-band outnumbers significantly the results, available for B and
R bands, having approximately the similar accuracy. Therefore, it is reasonable to begin
the search for possible variability from the dataset of the V-band photometry.

\subsection{Methods to search for light variability}
\label{ssec:stat.methods}

\citet{Sokolovsky17} presented the comparison of various methods to detect photometric variability
in time series data in the search for variable stars. We use in the following the two of their methods.

The first one is the well-known $\chi^{2}$ test. It allows to clearly define, based on the published
'critical' values of $\chi^{2}$ distribution and the number of observed data,
the confidence level, at which the Null hypothesis on the absence of variability can be rejected (assuming that
the measured data have Gaussian distribution). Here, statistics $\chi^{2}$ is
$\Sigma (x_{\rm i} - m_{\rm w})^2 /\sigma_{\rm i}^2$,  where $\sigma_{\rm i}^2$ are dispersions of
$x_{\rm i}$, assumed to have Gaussian distribution and
\mbox{$m_{\rm w} = \Sigma (x_{\rm i}/\sigma_{\rm i}^2$)/$\Sigma (1/\sigma_{\rm i}^2$)} is
the weighted mean of $x_{\rm i}$.

The second method uses the so-called robust median statistic (RoMS, $\eta$) \citep{Enoch03, Rose07}.
This is defined as
\mbox{$\eta$ = $\Sigma |x_{\rm i} - median(x_{\rm i})|/\sigma_{\rm i}$}.
The reduced RoMS, $\bar{\eta}$ with N--1 degrees of freedom, is defined as follows: \\
\mbox{$\bar{\eta}$ = (N--1)$^{-1} \times$ $\Sigma |x_{\rm i} - median(x_{\rm i})|/\sigma_{\rm i}$}.

For this method, there is no published theoretical distribution and respective 'critical' values for various
confidence levels.
As \citet{Enoch03} note, if the data $x_{\rm i}$ represent intrinsically the constant value, the expected value
of $\bar{\eta}$ is less then one. The value of $\bar{\eta}$ larger than 1 indicates probable variability. To quantify the
likelihood of variability, the authors ran a thousand of Monte Carlo simulations for data samples representing the
 constant value with the random noise. They adopt the values of $\bar{\eta}_{\mathrm 95}$, $\bar{\eta}_{\mathrm 98}$
and $\bar{\eta}_{\mathrm 99}$, for which 95, 98 and 99~per cent of realisations had $\bar{\eta}$ less than these
'critical' values of $\bar{\eta}$.
From their Table~2, it follows that the 'critical' value $\bar{\eta} = $ 1.2 was exceeded only in 10 cases of 1000.
Hence, one can treat this level of $\bar{\eta}$ as the critical point to reject the Null hypothesis on 'non-variable signal'
at the confidence level of $\alpha$ = 0.01, or, in the other words, the evidence of detecting variability with probability
of P = 1 -- $\alpha$ = 0.99.
The method was further successfully tested on the massive arrays of data by \citet{Rose07} and \citet{Burdanov14}.
While for the Gaussian distribution of errors, the RoMS method seems to be less powerful, than $\chi^{2}$ test
(in the sense of the lower confidence probability for a given light curve), this can serve as an independent
check of the suspected variability.

\subsection{Results of checks of possible variability}
\label{ssec:var.results}

In the top part of Table~\ref{tab:stat}, for Knots 1, 2, 4, 5, 6 we present the estimates of $\chi^{2}$
and the related probability of variability  along with the similar data for the reduced RoMS  $\bar{\eta}$. Here,
the estimates are given for the nominal errors in the Tables with the results of photometry for individual Knots.
A brief discussion of the possible effect of the underestimation of the observational errors is presented at
the end of this Section.

In Table~\ref{tab:stat}, we present
the ranges of the magnitudes $m_{\rm i}$ (excluding the low-accuracy points on the edges of the range),
the weighted mean magnitudes $m_{\rm w}$, the weighted mean error
\mbox{$\sigma_{\rm w}$ = $\sqrt(N)$/$\Sigma$ (1/$\sigma_{\rm i}^2$)},
the weighted rms$_{\rm w}$ around the weighted mean, and the related value of
\mbox{$\chi^{2}$ = $\Sigma (x_{\rm i} - m_{\rm w})^2 /\sigma_{\rm i}^2$}, the $median$ of all $m_{\rm i}$, and
the reduced parameter RoMS,
\mbox{$\bar{\eta}$  = $\Sigma$ ($| m_{\rm i} - median |$)/$\sigma_{\rm i}$)/(N--1)}.
For all Knots, the number of  observations in V-band is N = 43.

In the last two columns, we present the probability \mbox{P = 1 -- $\alpha$} for rejection of the Null hypothesis on the
non-variability of the examined samples according to criterion of $\chi^{2}$ and the
value of $\bar{\eta}$. For $\chi^{2}$ test this is based on the critical points of the $\chi^{2}$ distribution
for the respective values of probability, from Table~2.2a in the book by \citet{Bol'shev1983}.
As for probabilities, based on the statistics $\bar{\eta}$, we adopt the results of modelling, presented
in \citet{Enoch03}.

As one can see in the top part of Table~\ref{tab:stat}, the apparent amplitude of light variations ranges from 0.08
(Knot~5) to 0.16~mag (Knot~6), with the intermediate amplitudes of 0.09, 0.13 and 0.11~mag for Knots 1, 2 and 4.
The weighted rms varies between $\sim$0.016 and $\sim$0.046~mag for various Knots. For the nominal values of the measurement
errors, both statistics, $\chi^{2}$ and $\bar{\eta}$, appear sufficiently large for all 5 Knots, so that the confidence
levels $\alpha$ to reject the Null hypothesis, are 0.0005 for $\chi^{2}$ and 0.01 for $\bar{\eta}$.

While this result looks somewhat surprising, it is worth  of commenting the issue of how robust are the conclusions
on the variability of the considered Knots.
We adopt that the observed light variations reflect the real processes, that is the internal variations plus the noise
signal inherent to the measurement methodics. So, the characteristic measure of the significance of variations in both
tests, $\chi^{2}$ and $\bar{\eta}$, is the averaged ratio of the characteristic amplitude of variations to the measurement
error.

Therefore, if we, due to the incomplete understanding of the nature of the measurement errors, underestimate
them, we get, as a result, the elevated values of both $\chi^{2}$ and $\bar{\eta}$, and, in turn, the higher
confidence levels for detection of variability. In order to check, how the conclusions about the probable variability are
conservative, we vary upward all the
measurement errors by the factors of 1.41 and 2.0. The former case corresponds roughly to the situation,
in which we count the Poisson noise within the Knot aperture twice. We summarise the effect of the underestimated
error by a factor of 2.0 in the bottom part of Table~\ref{tab:stat}.  For the intermediate factor of 1.41,
the probability of variability, based on $\chi^{2}$, P $>$ 0.9995 for all Knots but Knot~5, for which this is 0.995.
Respectively, the probability of variability, based on $\bar{\eta}$ is P$\geq$0.99 for all Knots but Knot~5, for which
it is 0.90.

As one can see, for the underestimation factor of 1.41, the probability of rejecting the Null
hypothesis drops substantially only for Knot~5. For the other four Knots, the variability manifests  with the high confidence.
Even for the case of the underestimation factor of 2, the results on the variability of Knots 4 and 6 remain robust, for
Knot~2, the variability gets less confident, while the variations of light in Knots~1 and 5 appear non significant.

\begin{table*}
\begin{center}
\caption{Statistical parameters of variability for Knots 1, 2, 4, 5, 6}
\label{tab:stat}
\begin{tabular}{l|r|l|l|l|r|l|l|r|r}\hline  \hline \\ [-0.2cm]
\MC{1}{c|}{Name} &
\MC{1}{c|}{V~range}&
\MC{1}{c|}{V$_{\rm mean,w}$}&
\MC{1}{c|}{$\sigma_{\rm w}$}&
\MC{1}{c|}{rms$_{\rm w}$} &
\MC{1}{c|}{$\chi^{2}$}     &
\MC{1}{c|}{V$_{\rm med}$}    &
\MC{1}{c|}{$\bar{\eta}$}     &
\MC{1}{c|}{Prob.~var.}       &
\MC{1}{c|}{Prob.~var.}     \\

\MC{1}{c|}{ } &
\MC{1}{c|}{mag}&
\MC{1}{c|}{mag} &
\MC{1}{c|}{mag} &
\MC{1}{c|}{mag} &
\MC{1}{c|}{} &
\MC{1}{c|}{mag} &
\MC{1}{c|}{} &
\MC{1}{c|}{on $\chi^{2}$} &
\MC{1}{c|}{on $\bar{\eta}$} \\

\MC{1}{c|}{(1)} &
\MC{1}{c|}{(2)} &
\MC{1}{c|}{(3)} &
\MC{1}{c|}{(4)} &
\MC{1}{c|}{(5)} &
\MC{1}{c|}{(6)} &
\MC{1}{c|}{(7)} &
\MC{1}{c|}{(8)} &
\MC{1}{c|}{(9)} &
\MC{1}{c|}{(10)} \\
\\[-0.2cm] \hline \\[-0.2cm]
\multicolumn{10}{c}{ For case of the nominal errors. } \\
\hline
 Knot~1&  18.37-18.46 & 18.413 &0.009   & 0.021  & 229  &18.41 & 1.70 & $>$0.9995 & $>$0.99   \\ %
 Knot~2&  19.56-19.69 & 19.622 &0.016   & 0.040  & 270  &19.61 & 2.02 & $>$0.9995 & $>$0.99   \\ %
 Knot~4&  18.67-18.78 & 18.713 &0.012   & 0.029  & 290  &18.72 & 2.05 & $>$0.9995 & $>$0.99   \\ %
 Knot~5&  19.08-19.16 & 19.103 &0.009   & 0.016  & 151  &19.10 & 1.47 & $>$0.9995 & $>$0.99   \\ %
 Knot~6&  19.36-19.52 & 19.441 &0.012   & 0.046  & 641  &19.44 & 2.85 & $>$0.9995 & $>$0.99   \\ %
\multicolumn{10}{c}{For case of errors larger than nominal by a factor of 2.0.} \\
\hline \\[-0.2cm]
 Knot~1&  18.37-18.46 & 18.413 &0.009   & 0.021  &  57  &18.41 & 0.85 & 0.90      & ...       \\ %
 Knot~2&  19.56-19.69 & 19.622 &0.016   & 0.040  &  67  &19.61 & 1.01 & 0.99      & 0.90      \\ %
 Knot~4&  18.67-18.78 & 18.713 &0.012   & 0.029  &  72  &18.72 & 1.03 & 0.995     & 0.90      \\ %
 Knot~5&  19.08-19.16 & 19.103 &0.009   & 0.016  &  38  &19.10 & 0.74 & ...       & ...       \\ %
 Knot~6&  19.36-19.52 & 19.441 &0.012   & 0.046  & 160  &19.44 & 1.43 & $>$0.9995 & $>$0.99   \\ %
\hline \\

\multicolumn{10}{l}{Col.~2: full range of variations in $V$ band; Col.~3: weighted mean V-mag; Col.~4: weighted mean error } \\
\multicolumn{10}{l}{$\sqrt(N)$/$\Sigma$ (1/$\sigma_{\rm i}^2$); Col.~5: weighted rms about the weighted mean; Col.~6: Value of $\chi^{2}$; Col.~7: median} \\
\multicolumn{10}{l}{V-mag; Col.~8: value of parameter $\bar{\eta}$; Col.~9: Probability of 'variability', P = 1 -- $\alpha$, based on $\chi^{2}$;} \\
\multicolumn{10}{l}{Col.10: similar probability, based on $\bar{\eta}$. See Sect.~\ref{sec:vary} and \ref{ssec:var.others}. For all Knots, N$_{\rm obs}$ = 43.} \\
\end{tabular}
\end{center}
\end{table*}

\begin{figure*}
  \centering
 \includegraphics[angle=-90,width=7.5cm,clip=]{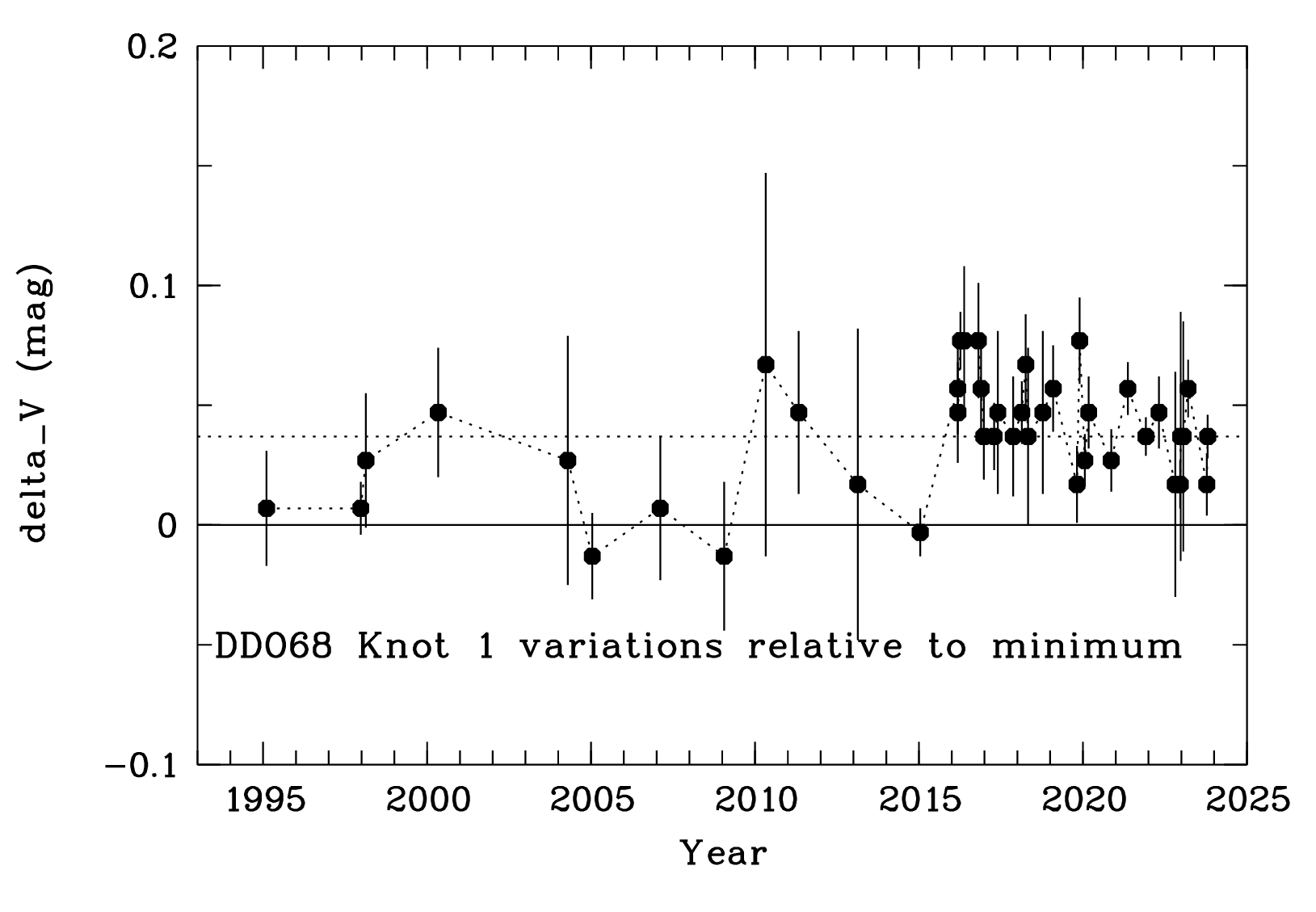}
 \includegraphics[angle=-90,width=7.5cm,clip=]{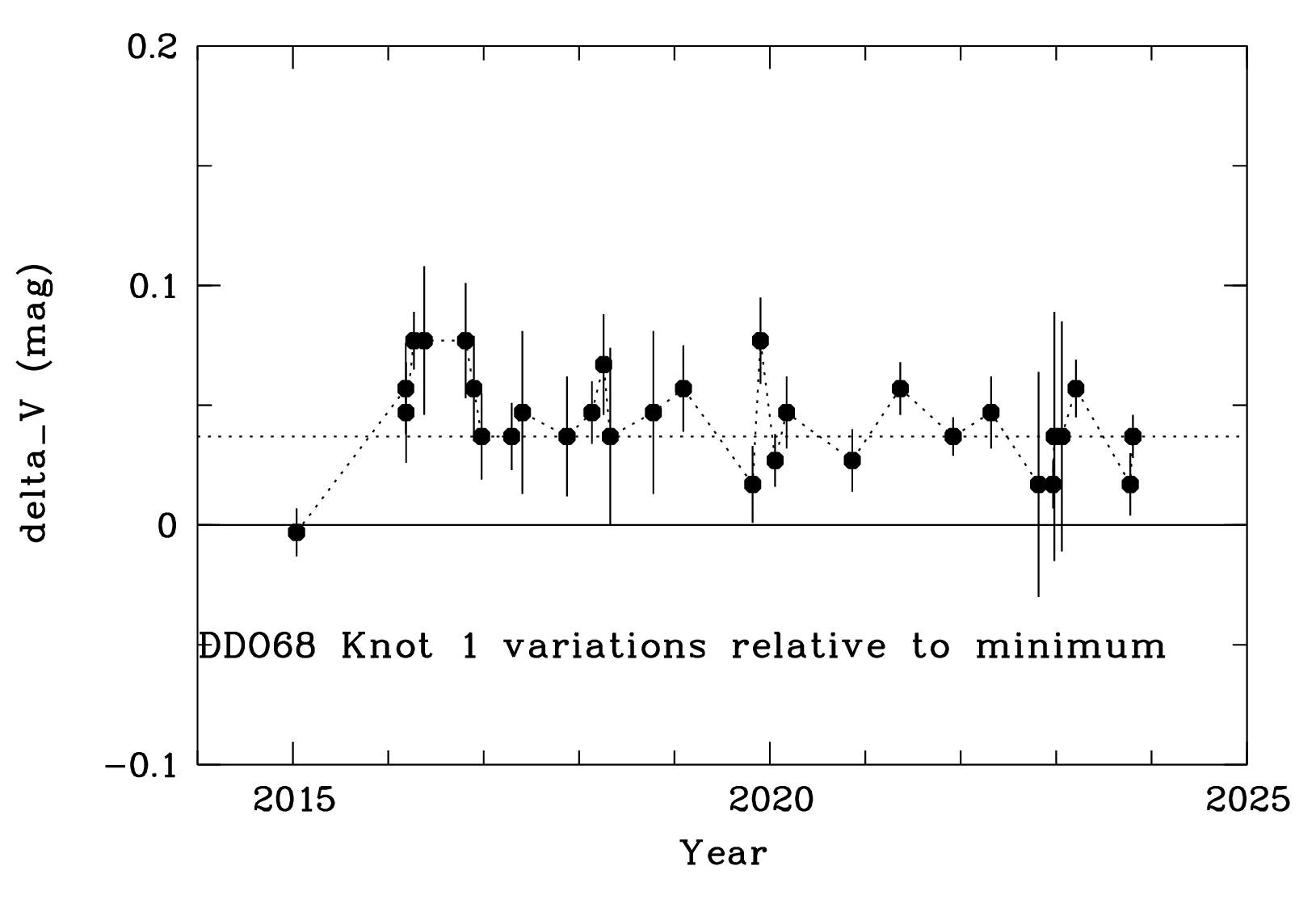}
  \caption{
Light curves of 'variable' component of Knot 1 (relative to its minimal brightness of V=18.447~mag)
in V-band in two time intervals: 1995--2023
and 2015--2023.
}
  \label{fig:lcurves_K1_var}
 \end{figure*}

\begin{figure*}
  \centering
 \includegraphics[angle=-90,width=7.5cm,clip=]{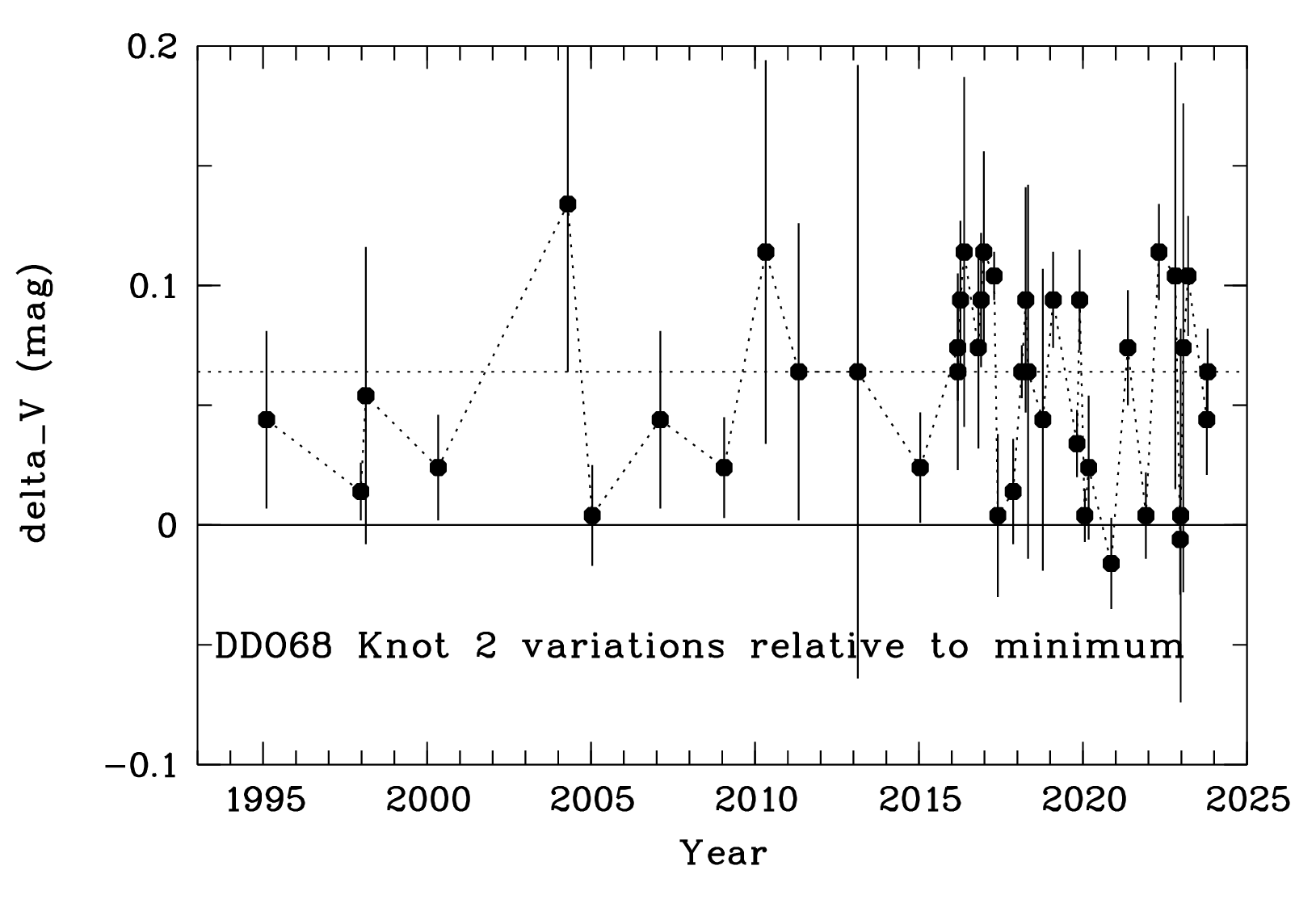}
 \includegraphics[angle=-90,width=7.5cm,clip=]{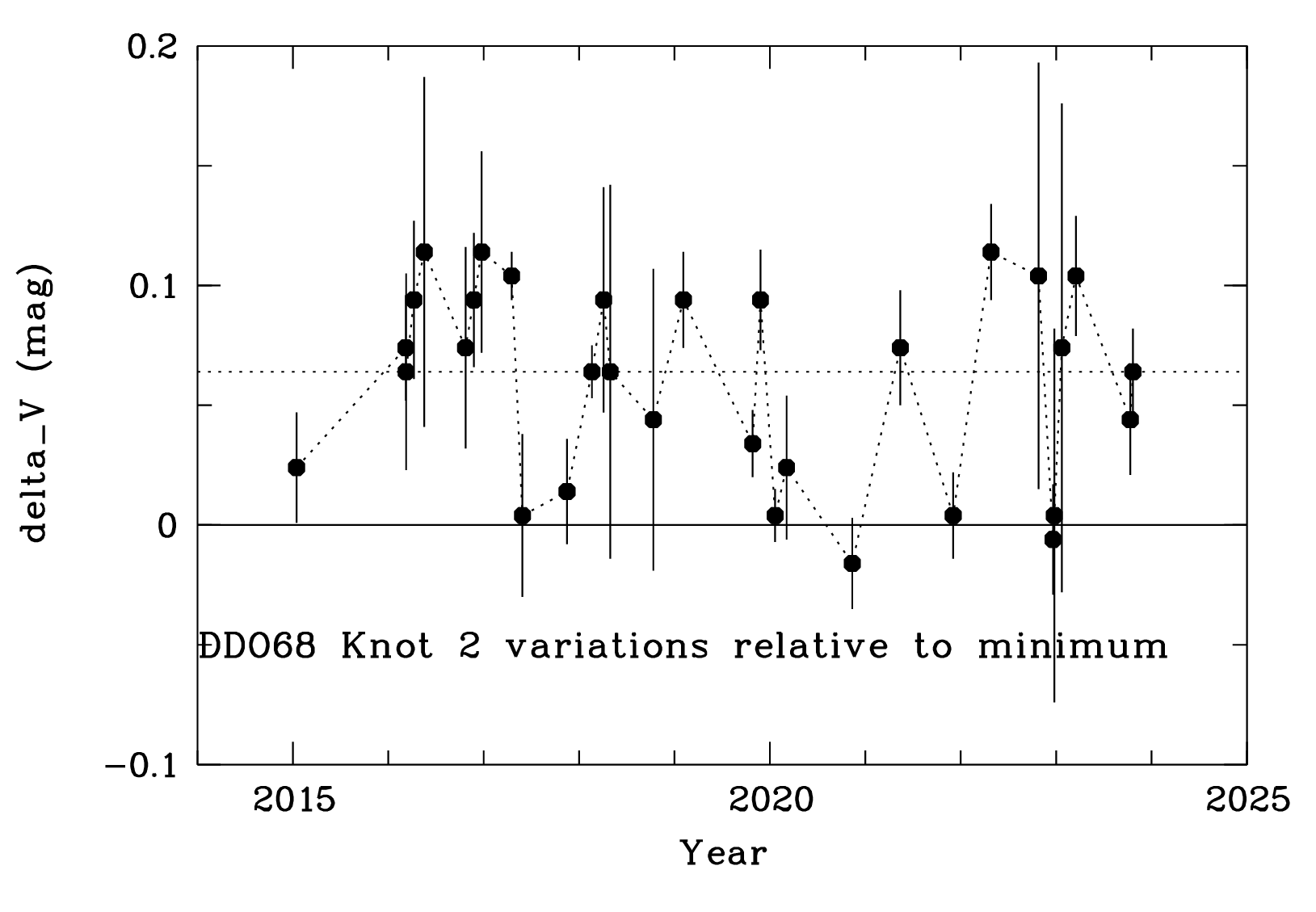}
  \caption{
Light curves of 'variable' component of Knot 2 (relative to its minimal brightness of V=19.674~mag)
in V-band in two time intervals: 1995-2023
and 2015-2023.
}
  \label{fig:lcurves_K2_var}
 \end{figure*}

\subsection{Lightcurves  of the residual components}
\label{ssec:resid.lightcurves}

It is instructive to see the light variations in the analysed Knots at the higher contrast,
after the underlying non-variable component light is subtracted. This can give us indications on the
variability amplitudes and timescales of the respective objects and to hint the type of the related stars.

\begin{figure*}
  \centering
 \includegraphics[angle=-90,width=7.5cm,clip=]{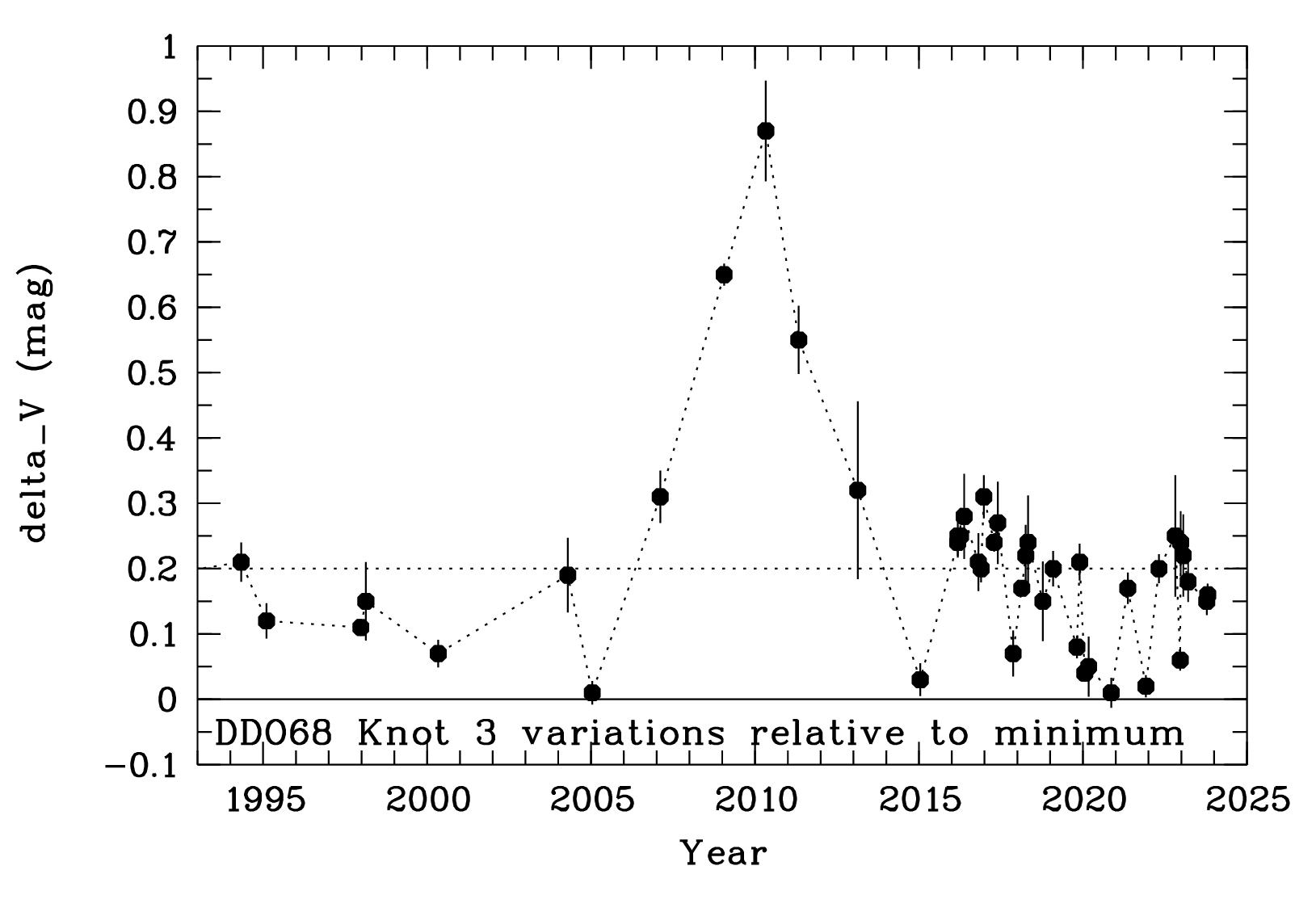}
 \includegraphics[angle=-90,width=7.5cm,clip=]{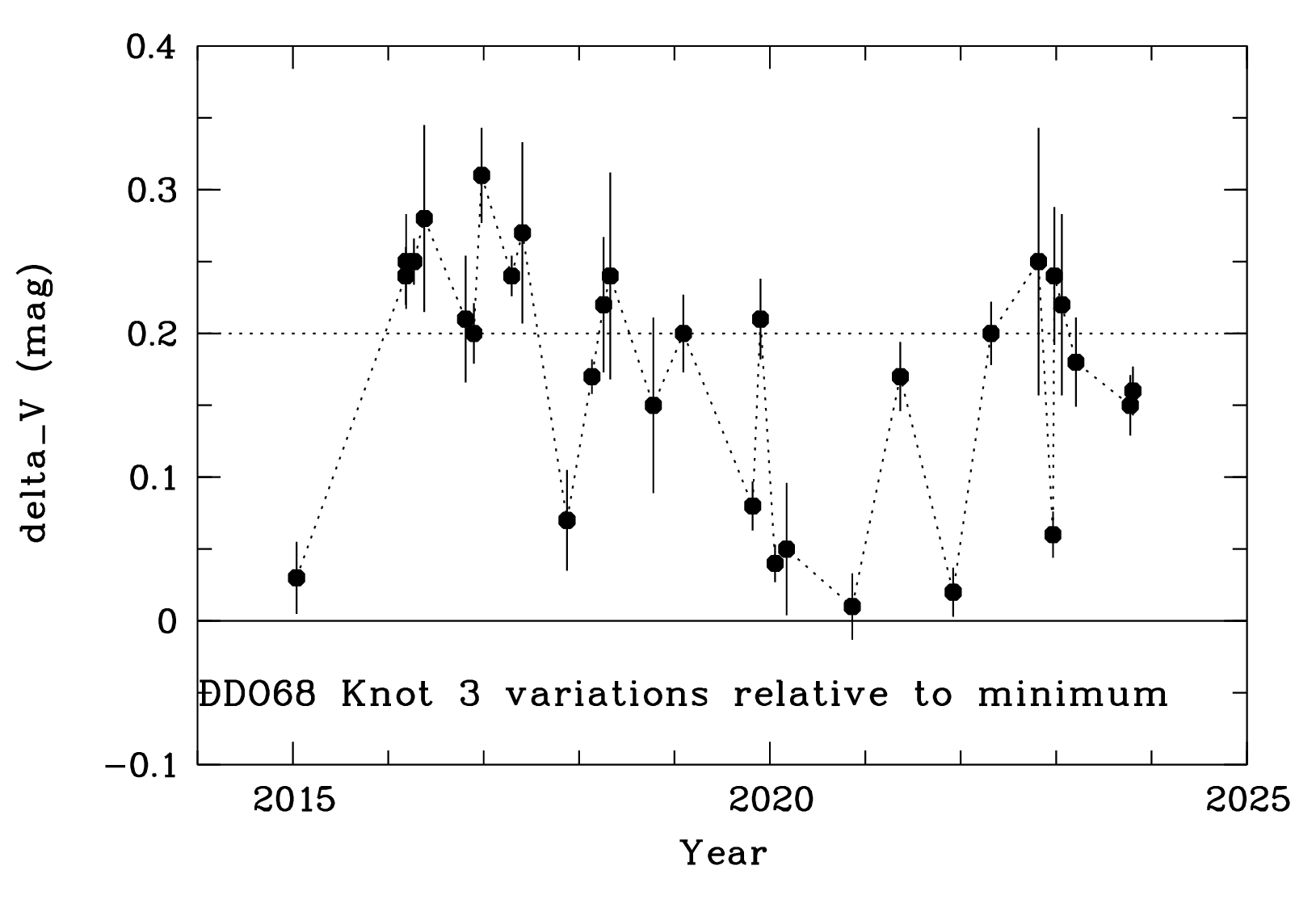}
  \caption{
Light curves of 'variable' component of Knot 3 (relative to its minimal brightness of V=20.20~mag)
in V-band in two time intervals: 1995-2023
and 2015-2023.
}
  \label{fig:lcurves_K3_var}
 \end{figure*}

\begin{figure*}
  \centering
 \includegraphics[angle=-90,width=7.5cm,clip=]{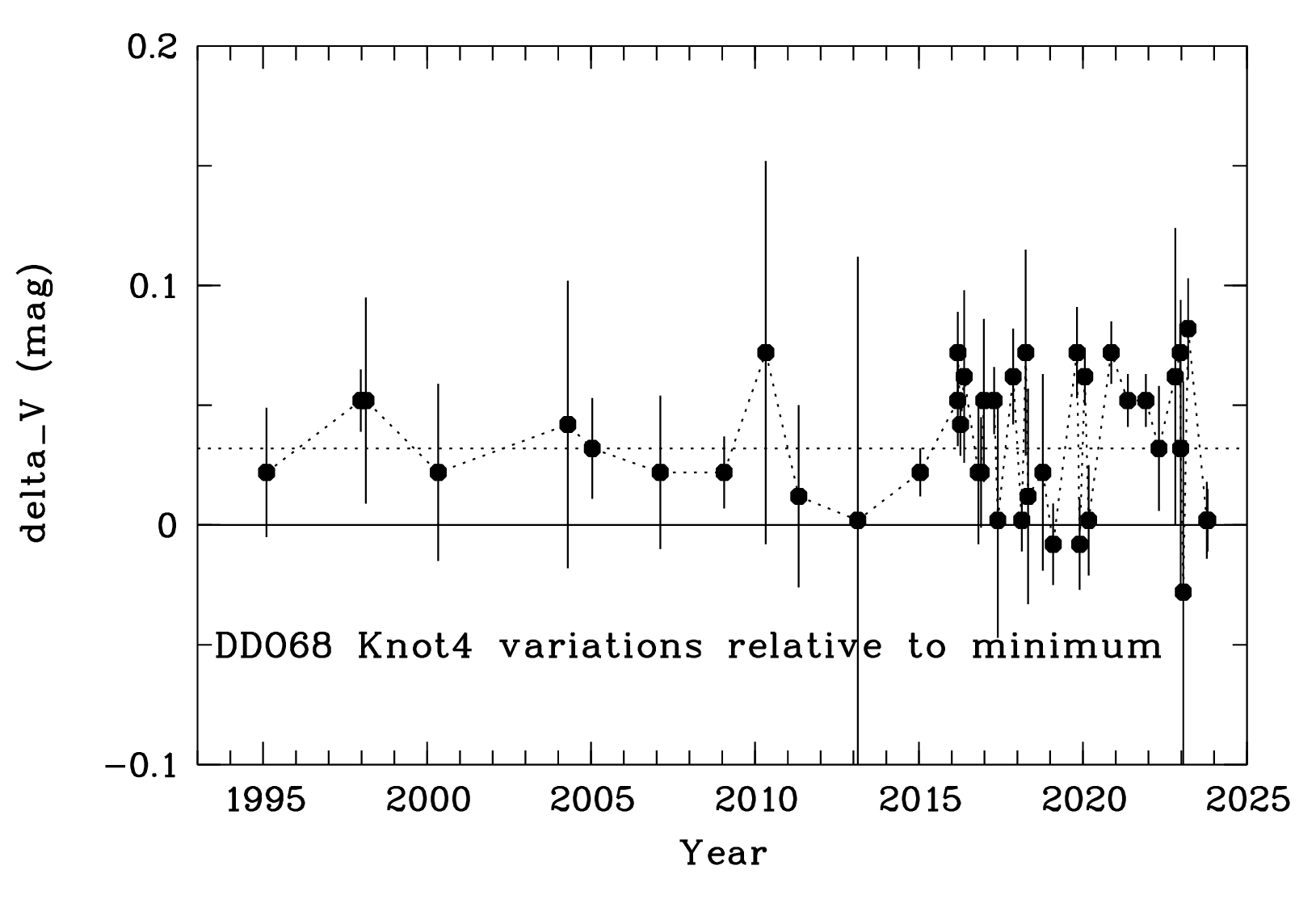}
 \includegraphics[angle=-90,width=7.5cm,clip=]{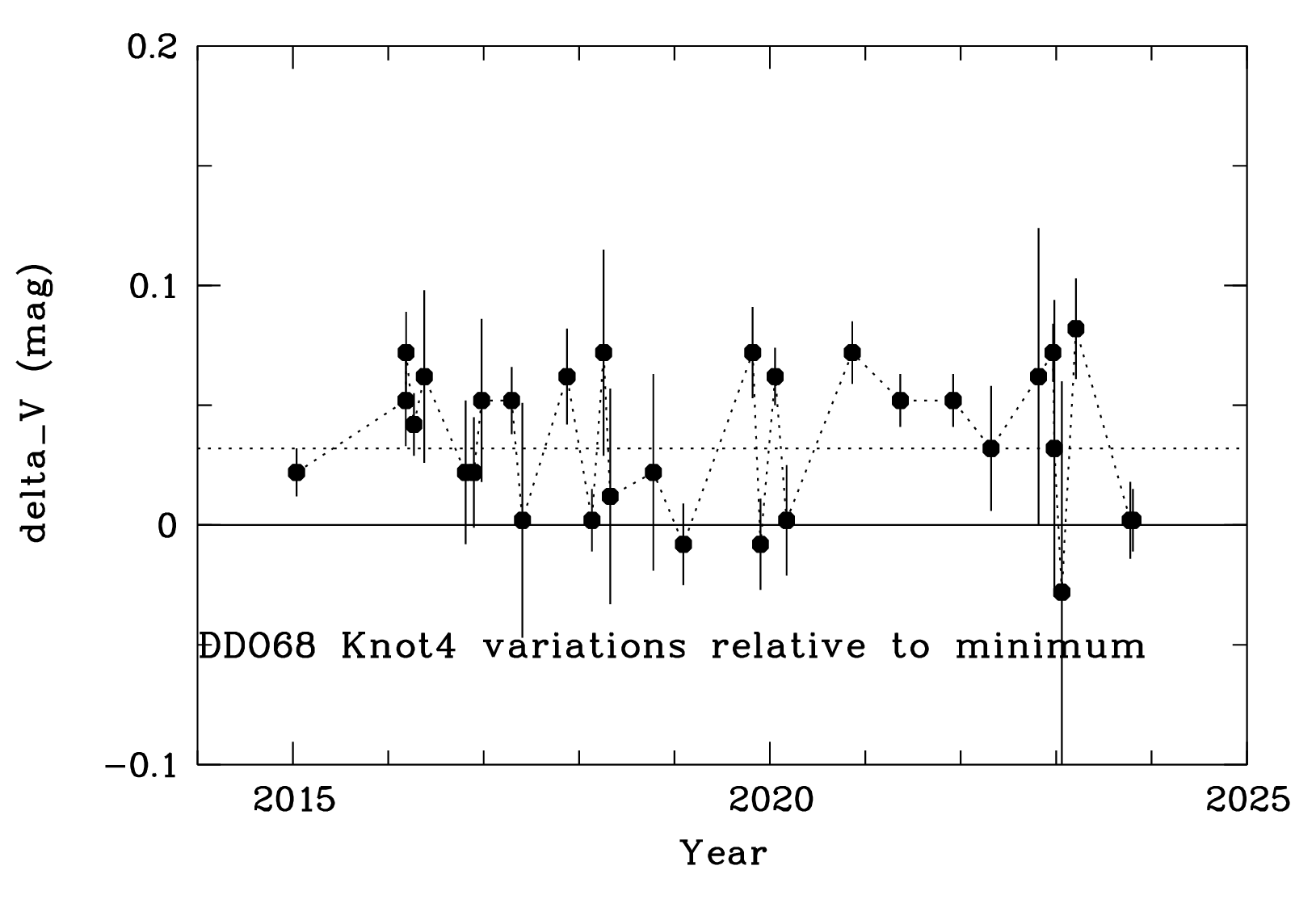}
  \caption{
Light curves of 'variable' component of Knot 4 (relative to its minimal brightness of V=18.775~mag)
in V-band in two time intervals: 1995-2023
and 2015-2023.
}
  \label{fig:lcurves_K4_var}
 \end{figure*}

\begin{figure*}
  \centering
 \includegraphics[angle=-90,width=7.5cm,clip=]{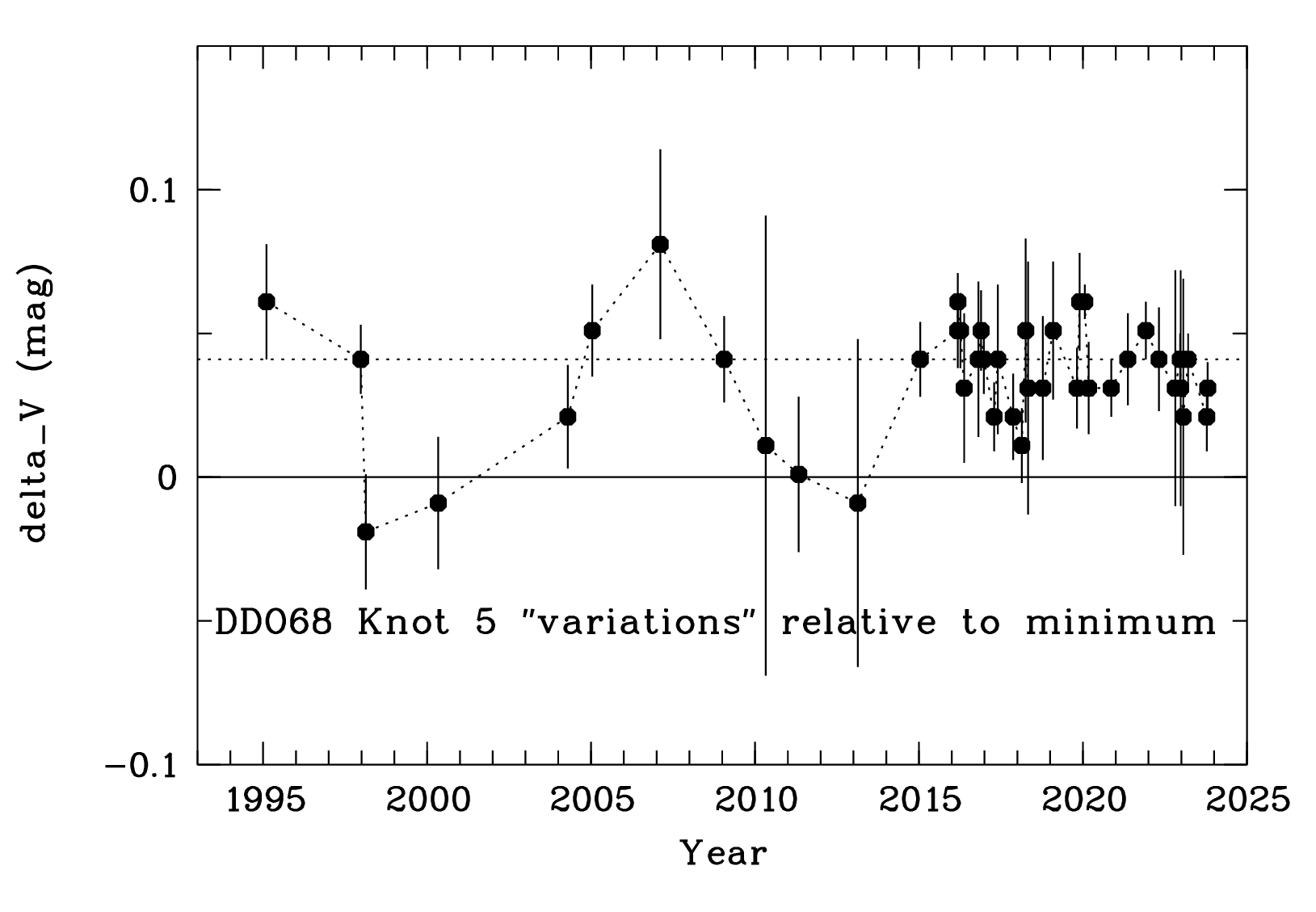}
 \includegraphics[angle=-90,width=7.5cm,clip=]{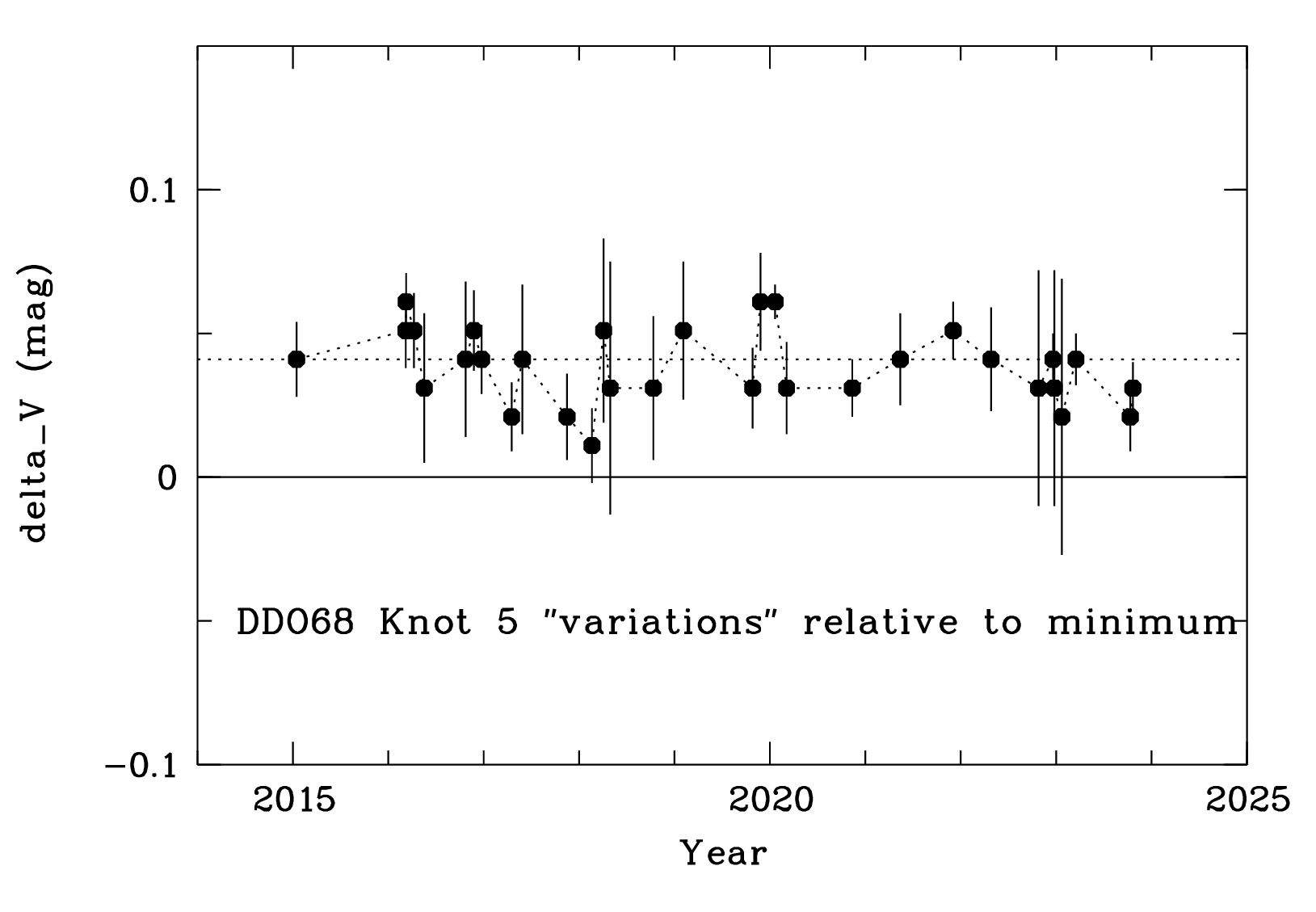}
  \caption{
Light curves of 'variable' component of Knot 5 (relative to its minimal brightness of V=19.141~mag)
in V-band in two time intervals: 1995-2023
and 2015-2023.
}
  \label{fig:lcurves_K5_var}
 \end{figure*}

\begin{figure*}
  \centering
 \includegraphics[angle=-90,width=7.5cm,clip=]{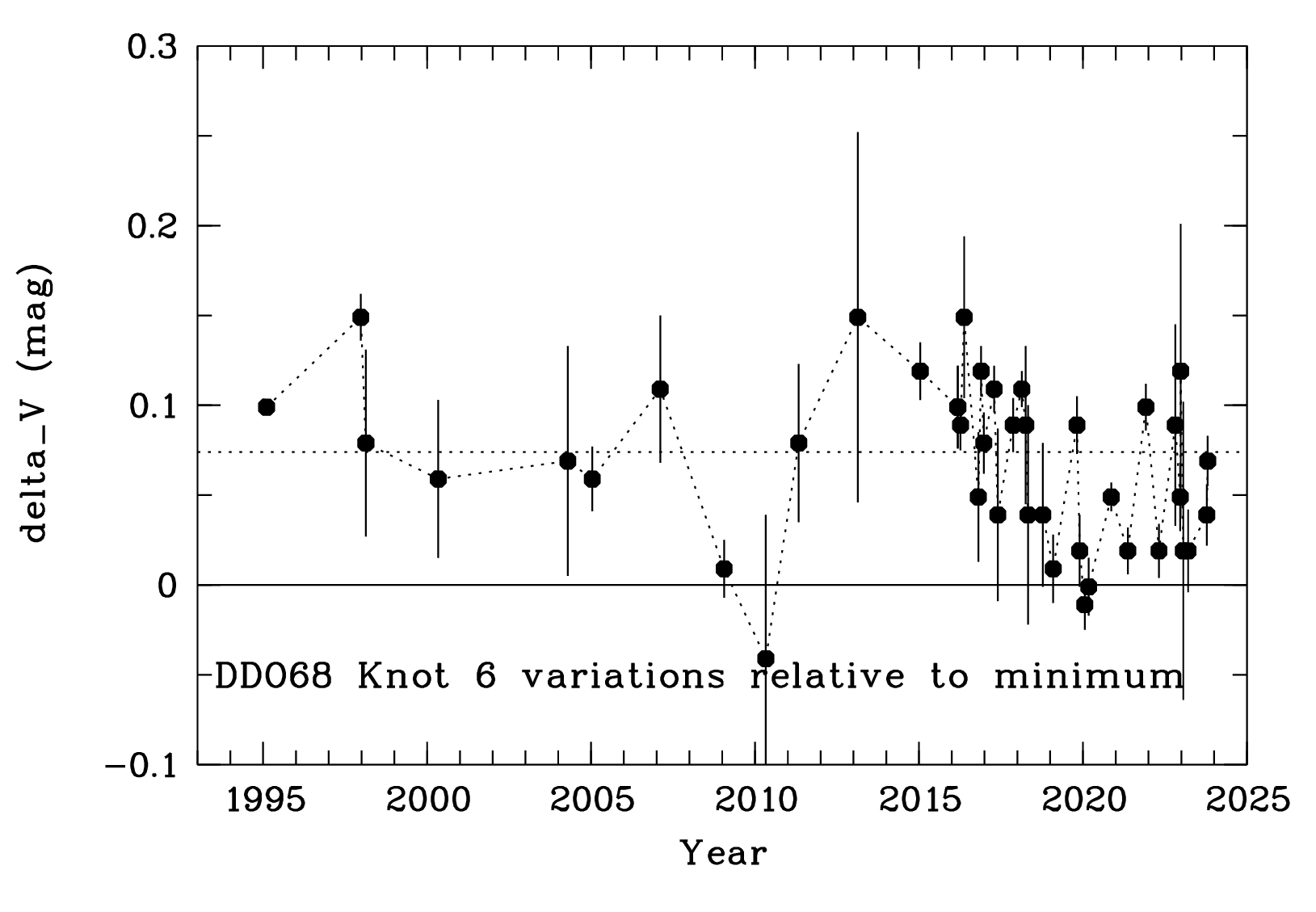}
 \includegraphics[angle=-90,width=7.5cm,clip=]{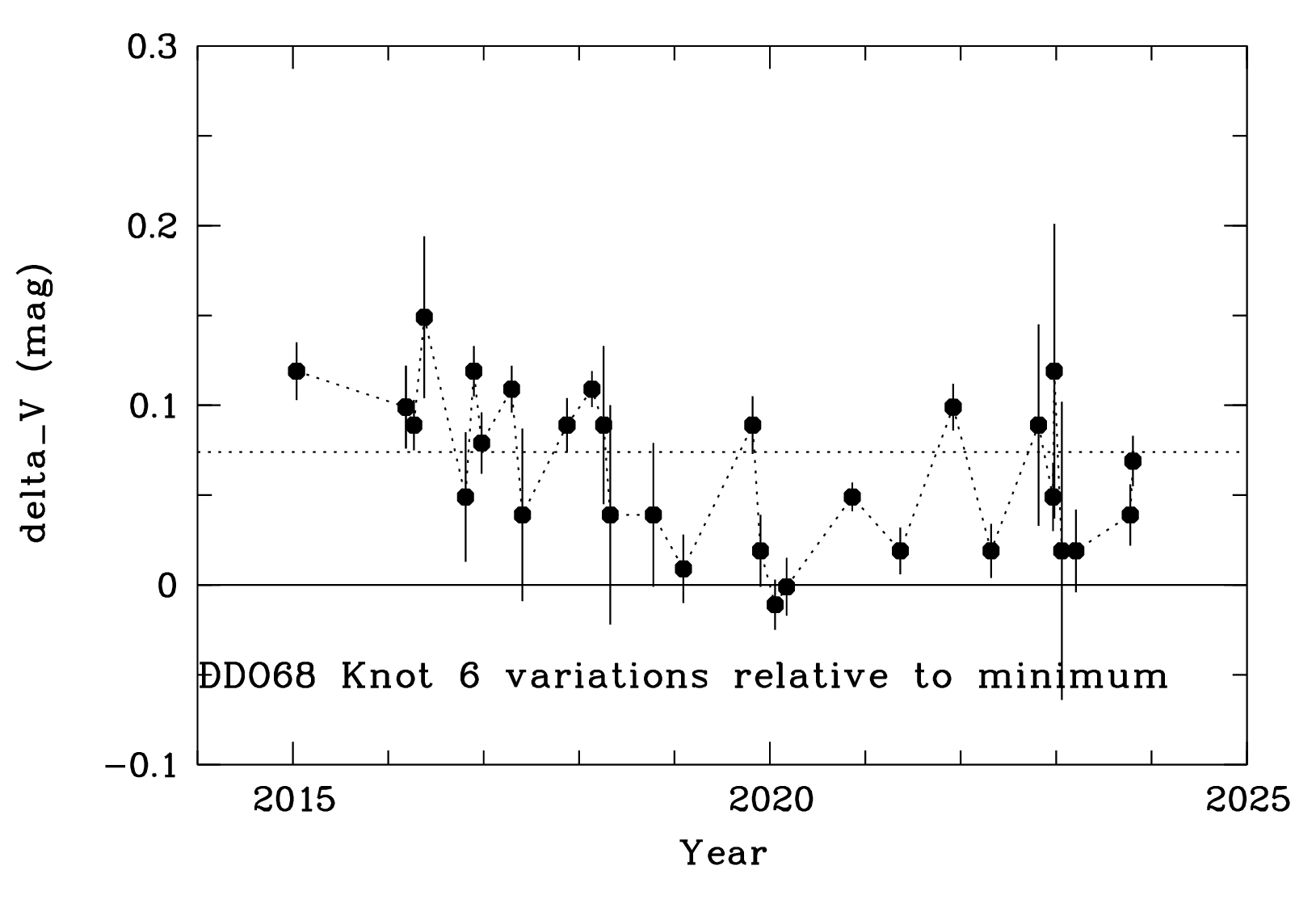}
  \caption{
Light curves of 'variable' component of Knot 6 (relative to its minimal brightness of V=19.509~mag)
in V-band in three time intervals: 1995-2023 
and 2015-2023.
}
  \label{fig:lcurves_K6_var}
 \end{figure*}

For all Knots (apart Knot~3),  we estimated the level of the underlying background, taking several
points with the minimal brightness and estimating their weighted mean and the scatter around this.
The  adopted  minimal V-band magnitudes are given in the captions of the respective plots in
Figures~\ref{fig:lcurves_K1_var}--\ref{fig:lcurves_K6_var}.
These lightcurves are shown in two time intervals (years 1995--2023 and 2015--2023)
in order to better distinguish visually variations at the longer and the shorter timescales.
The minimal brightness is shown in all light curves as the solid line at $\delta$V = 0. We draw additionally
for each plot the dotted line, corresponding to the median magnitude over the whole period.

\section[]{DISCUSSION}
\label{sec:dis}

\subsection{Variability of DDO68-V1}

The variability of LBV in DDO68 (DDO68-V1 in Knot~3) was already discussed
in many papers, including the variability of spectral features and variations of its
optical luminosity \citep[e.g.,][]{LBV, DDO68LBV, IT09, LBV_giant, Guseva2022}.
Its main features are
the absence in emission of any metal lines while the observed Balmer lines of Hydrogen
and the lines of He~I show P~Cygni profiles with the typical radial velocities of $\gtrsim$800~\kms.
As for the luminosity variations, this star  was caught in the phase of the giant eruption
during the years 2008--2011, when its luminosity in V-band approached M$_{\rm V}$ $\sim$ --10.6~mag.
The total amplitude of its variability was estimated of $\sim$4~mag.
Since before the LBV giant eruption, there were no more or less regular observations of this object,
our presented  monitoring was intended, in particular, to examine the behaviour of the LBV light curve
after such an event.

  The spectral monitoring of Knot~3 in 12 epochs during years 2008--2018 was presented
recently by \citet{Guseva2022}. In particular, they discovered variations of the LBV broad H$\alpha$-line
flux by a factor of a thousand during their decade-long observations.
The two latest their measurements, in April 2016 and April 2018, are obtained
within several days from our imaging observations. In April 2016, the flux of the broad H$\alpha$-line
was 20 times lower than its maximal value near the peak of the 'giant eruption, while in April 2018 it was
below the detection level, that is at least several times smaller.
Our light curve of Knot~3 at these epochs (see Table~\ref{tab:photo_Kn3} and Figure~\ref{fig:lcurves_K3_var}),
is consistent with the fall of the LBV light after the giant eruption to the deep minimum in 2015,
with the subsequent episodic maxima mimicking the S~Dor phase variations. The latter spectral observations of the LBV
appear close to these local maxima. The behaviour of the broad H$\alpha$-line flux indicates that this is the tracer
of the strongly decaying 'giant eruption' shell. 
We postpone the description and analysis of the Knot~3 data and the derived light curves for DDO68-V1 to
the forthcoming dedicated paper.

\subsection{Variability of the other star-forming regions}
\label{ssec:var.others}

We use all available DDO68 images to check the potential of the
integrated photometry of the most metal-poor SF regions to search
for new variable supergiants with the lowest known metallicities. This variant was
suggested in the paper of \citet{DDO68LBV} for the ground-based
telescopes situated in observatories with the superb seeings. Of course, the better
seeing in images of this kind monitoring programs, the simpler one can separate
individual variable supergiants.
In fact, our results described below, suggest that this task can be
partly accomplished with more relaxed conditions if the accuracy of photometry,
the cadence period and the time span of monitoring are optimised for the expected parameters
of the searched variability.

For Knot~1, the results of our statistical tests are indicative on its V-band variations
in the range 18.37--18.46~mag, with the median of V$_{\rm Kn.1,med}$ = 18.41.
As one can see in Table~\ref{tab:stat}, the detection of variability is confident, but
somewhat less robust with
respect of the (potentially) underestimated measurement errors in comparison to Knots 6, 4 and 2.
We conduct the preliminary consideration
of possible stars that can be responsible for the observed amplitude of light variations of
$\delta V_{\rm Kn.1}$ $\sim$ 0.09~mag. The two brightest stars
of this region, No.~13 and No.~18 according to the list in \citet{DDO68LBV}, are blue
supergiants (BSG) with the measured on May 2, 2010 V-band magnitudes of 23.81 and 24.13~mag,
and the colour (V--I) = --0.09 and --0.21~mag, respectively.
The corresponding absolute magnitudes are M$_{\rm V}$ = --6.77 and --6.45~mag.
The V-band light of star No.~13 is $\sim$5.4~mag fainter than the total light of Knot~1,
that is comprises only 0.7~per~cent of this. The star No.~18 contributes
to the total V-band light of Knot~1 only 0.5~per~cent. 
The observed light variations of Knot~1
 imply very strong variations in one or both of these two blue supergiants:
$\delta V_{\rm No.13}$ $\sim$3.0~mag, and $\delta V_{\rm No.18}$ $\sim$3.3~mag.

  Such large amplitudes are quite atypical for blue supergiants. Even LBVs during the so-called
S~Dor stage (normal eruptions) show amplitudes at the time-scale of years of up to 2.5~mag \citep[e.g.][]{vanGenderen01}.
Only for a few observed Milky Way LBVs, such as $\eta$~Car, P~Cyg and a number of extragalactic LBVs,
such as NGC 2363-V1 \citep{Petit06}, LMC R127 \citep{Walborn08} and UGC2773-OT \citep{Smith2016}, the so-called
'giant eruptions' are registered, with the amplitudes of four and more magnitudes. Similarly, the large amplitude
events are observed for a number of the so-called Supernova impostors. However, their relation to a certain
type of massive stars is not yet clear.

As for the data
on the S~Dor phase amplitudes up to $\sim$2.5~mag, they all are picked up on the sample of relatively
metal-rich LBVs in the Milky Way, Magellanic Clouds, M31 and other 'metal-rich' galaxies. Therefore, they can
be not representative of LBVs in the XMP regime.

In fact, the S~Dor phase amplitudes may depend on the LBV metallicity. The example of DDO68-V1 directly
indicates to this possibility. Of course, the single unique case
of variability of DDO68-V1 does not pretend to have statistical implications.  However, the precedent of its
S~Dor variations with the amplitude of up to 3.5~mag, discovered in our forthcoming paper, gives the clear
hints on the probable larger S~Dor type variations in the XMP LBVs. However, to be confident that these variations
are consistent with the typical S~Dor behaviour, that is of almost constant bolometric luminosity and the related
changes of colours versus the luminosity \citep[e.g., ][]{Solovyeva2019},
it takes to analyse the multi-colour data for a wide range of DDO68-V1 light variations.

Finally, we should comment on two rather bright non star-like objects near the center of Knot~1, with
V $\sim$ 21.3--21.5~mag.
These can be compact \HII\ regions, excited by the hot massive stars. Their contribution to the total light of Knot~1
can be an order of magnitude larger than that of the mentioned above BSGs (No.~13 and 18). Hence, their required
amplitudes of variability will be of less than $\sim$1~mag. However, such objects are too luminous for O-stars
at the main sequence. Hence, one expects that these \HII\ regions are excited by a 'cluster' of massive stars.
Therefore, they are less probable contributors to the observed variability of Knot~1.

For Knot~2, the available data in V-band indicate the range of
variations from 19.69 to 19.56 mag. Hence, one can suspect variability with
the amplitudes of $\sim$0.13 mag. The R-band data are much more sparse,
but indicate the similar amplitude of $\sim$0.10~mag, from 19.46 to 19.36~mag.
The B-band lightcurve is even more sparse, and due to the larger errors in the
extremums, there is no indication on the real variations larger than 0.05~mag.
The two brightest stars of this region are the blue supergiant No.~19 with
V = 24.13~mag and (V--I) = --0.21~mag, as on the epoch of the HST image (2010.05.02),
and the red supergiant No.~29 with V = 24.36~mag and (V--I) = 1.41~mag.
They contribute, respectively, 1.5 and 1.2 per cent of the total V-band light
of Knot~2. To produce the visible variations of 0.13~mag,
these stars should vary with the amplitude of $\sim$2.4--2.6~mag.

For Knot~4, the variations in V-band (18.67 to 18.78) indicate the amplitude of $\delta$V $\sim$ 0.11~mag.
There are about 10 blue and red supergiants within Knot~4, with V-mag from 23.00 to 24.40~mag.
They contribute to the integrated light of Knot~4 from $\sim$1.8 down to $\sim$0.3 per cent.
Therefore, for variations with the amplitude of $\sim$0.11~mag, the individual stars
should vary by a factor of 5 (1.8~mag) to 30 (3.7~mag). With that large number of
candidate strongly variable supergiants,  more data of high accuracy is necessary
for individual stars on the images of HST to try
to assign the visible variations of Knot~4 to a specific star or stars.

For Knot~5, the amplitude of V-band light variations does not exceed 0.08~mag.
Apart the young cluster itself, three nearby individual supergiant stars fall into the used aperture,
No. 7, 21 and 42, with V-band magnitudes of 23.3 to 24.5~mag. They contribute  to the total light within the used
aperture from $\sim$3.5 to $\sim$1 per cent. Therefore, the visible light variations in Knot~5 require
either variations of nearby stars by a factor of 2.5--8, or substantial variations of massive stars inside the cluster.

For Knot~6, the V-band variations are the most prominent, with the full range of
19.36 to 19.52~mag.
We also do see partly correlated variations in V and B bands over
the whole period of $\sim$18~years. Variations in R-band, despite the data
are significantly more sparse, also look like to occur in phase with those
in V-band. In Fig.~\ref{fig:lcurves_K4-K6} (bottom panel), showing all three B, V, R
lightcurves for Knot~6, we draw by the dotted lines the median values for
each band. We also draw by the solid line the estimated level of
of the minimal light in V band.

This allows us to estimate the full range of variations in V, B and R bands
of $\sim$0.16, $\sim$0.10, $\sim$0.10~mag, respectively. We first suggest that the observed
variability of Knot~6 is related to the light variations of its brightest supergiant, identified as
star No.~2 in the DDO68 list of the most luminous stars related to the
star-forming regions with the lowest gas O/H \citep{DDO68LBV}.
This candidate Yellow (warm) Hypergiant \citep[e.g.][]{Humphreys2013},
with the HST-based V = 21.47~mag and \mbox{M$_{\rm V}$ = --9.11~mag}, is the
second most luminous star after DDO68-V1 at that epoch (see Fig.~6 in \citet{DDO68LBV}).

At the HST image, the V-band light of this star comprises $\sim$20.5\%
of the rest Knot~6 light within the aperture with diameter of 6~arcsec. The brightness
increase by $\delta$V = 0.16~mag in the total light of Knot~6 corresponds to
the brightening of star No.~2 by 0.7~mag, reaching the absolute magnitude
of M$_{\rm V} \sim$ --9.8. According to its HST colour (V-I) = 0.33~mag, this
luminous star is currently classified as Yellow Hypergiant (YHG). While LBV stars spend most
of their life time as blue, their evolutionary tracks also reach the area of Yellow
Hypergiants, so one can find them occasionally in this phase \citep[e.g.][]{vanGenderen01}.

Besides, one can not exclude that its colour is affected by a circumstellar envelope, often observed
near evolved massive stars \citep[e.g.][]{vanGenderen01, Kniazev2017}. So that,
with the account of dereddening,  it can be bluer and belong to A, or even
to B hypergiants. Probably, the medium resolution spectroscopy of this star,
with the expected V-magnitude between 21.47 and $\sim$20.9~mag, will help to better
determine its spectral class.

One can also consider an alternative case, when the observed variability of Knot~6
is related to variability of the fainter blue supergiant, the star No.~11,
with M$_{\rm V}$ = --7.11~mag, and (V--I) = --0.11.
Its V-band magnitude of 23.42~mag, corresponds to $\sim$2.5~per~cent of
the integrated light of Knot~6. The other brightest stars situated around Knot~6,
No.~4, 10 and 17, are situated far outside the aperture with the diameter of 6~arcsec,
used to monitor the \HII\ region potential variability.
To produce the variations of the knot integrated light with
the amplitudes of 0.16~mag, this star should brighten by $\sim$7 times, or by $\sim$2.1~mag,
reaching M$_{\rm V}$ $\sim$ --8.2~mag. Thus, one can think on this B-supergiant
as a one more LBV candidate in the phase of S~Dor-type variability.

 From the examination of the lightcurves and tables with photometric data for the individual
DDO68 Knots, one can estimate the typical time scale of the significant variability (peak-to-peak)
in the Knots where the statistical analysis indicates a sufficiently high probability
of the real variations. These time scales can be useful for a more detailed analysis.

 In particular, for the strongest (after Knot~3) variability in Knot~6, one can see the
minima of the V-band light near epochs of 2009.1, 2020.0 and 2023.2,  while the maximal brightness
was reached near epochs of 1998.0, 2015.0--2016.3. That is the typical long-term variability of a star
in this Knot has the characteristic times of 2--5 years from bottom to top and back.
This is important to have in mind when checking the potential variability of the luminous
stars at the shorter timescales on the images available in the HST archive.
So, the estimated amplitudes of the brightest stars in these regions, of $\sim$0.7--3~mag
on the timescales of $\sim$2 years, will appear as only 0.03--0.12~mag variation on the timescale of 1~month.

The above analysis and discussion of the integrated light variations for various SF regions in DDO68,
implies that the detected amplitudes of 0.08--0.16~mag correspond to the brightening of their brightest supergiants
from the apparent V $\sim$ 23--24~mag (or V $\sim$ 21.5~mag for the YHG in Knot~6) to the levels of
V $\sim$ 20.5--21.5~mag. The stellar objects of such magnitudes
are accessible for the medium resolution spectroscopy with the ground-based large telescopes. While
their light contribution to the whole emission of the associated \HII-region comprises at most 10--15 per cent,
for the properly positioned long slit with the width of $\lesssim$1~arcsec, its contrast can increase several times.
Thus, the practical recommendation for the continuation of monitoring of the 'Northern Ring' includes
the spectral follow-up of the regions, in which one detects the bright phase of the lightcurve. This will allow  us
to pick-up the  preliminary information on the spectral properties of the most metal-poor supergiants/hypergiants
in their bright phases.

\section{Summary}
\label{sec:summ}

We summarise the presented results and the discussion above and draw the
following conclusions:

\begin{enumerate}
\item
The unique LBV  DDO68-V1 was discovered in January 2008 in
the void galaxy DDO68 near the center of \HII-region 'Knot~3' with the almost record-low
metallicity [12+$\log$(O/H) $\sim$ 7.1~dex]. The previous sparse lightcurve of Knot~3 for the
period of 2005--2015, when the LBV 'giant eruption' was discovered, was presented by \citet{DDO68LBV}.
Here we extend the lightcurve of Knot~3, adding our photometry with 3 telescopes for 35 epochs till the end
of 2023. We also use the archive data from 10 other telescopes. The light variations of Knot~3 with the
amplitude of up to $\sim$0.3~mag during the last 8 years are well documented.
The lightcurve for the LBV itself and its discussion will be presented in the forthcoming paper.
\item
 Apart Knot~3, our images cover five more SF regions (Knots) in or near the so-called DDO68
'Northern Ring'.
There are many various supergiant stars in these Knots, listed in \citet{DDO68LBV}, which can, in principle,
show large light variations. Such supergiant variability would manifest it as 'small' variations in the
total light of these Knots. We perform the photometry of these Knots, similar to that of Knot~3, and conduct
the statistical analysis of the respective data sets to check whether they are consistent
with the Null hypothesis on their non-variability.
\item
With the use of the well-known criterion $\chi^2$ and the additional statistics RoMS,
the Null hypothesis is rejected at the confidence level of $\alpha$ = 0.0005 (for $\chi^2$)
for Knots~1, 2, 4, 5, 6. Their peak-to-peak amplitudes are of $\delta$V $\sim$ 0.09, 0.13, 0.11, 0.08 and 0.16~mag,
respectively.
\item
The detected variations of Knots' light are naturally explained as manifestations of
variability in the brightest supergiants within these SF regions. Since the contribution of
individual supergiants to the integrated light of their Knots comprises of $\lesssim$1--10~per cent,
the related variations of the supergiants should reach amplitudes of $\delta$V $\sim$0.7--3.5~mag.
With so strong variations, one can expect to detect the additional LBV candidates in these Knots.
In Knot~6, if the light variations are related to those of the candidate warm hypergiant, their amplitude
can be only $\sim$0.7~mag.
\end{enumerate}

\section*{Acknowledgements}

The authors are grateful to A.~Valeev for help with BTA observations, to L.~van~Zee, D.~Hunter,
B.~Elmegreen, U.~Hopp,
L.~Makarova, R.~Swaters, B.~Mend\'ez, V.~Taylor, R.~Jansen, R.A.~Windhost,
S.C.~Odewan, J.E.~Hibbard  for providing archival CCD images of
DDO68, obtained for their observational programs. We are pleased to thank
P.~Kaigorodov and D.~Kolomeitsev for their kind help in extracting the data
from archive tapes. The authors thank the anonymous referee for careful reading the manuscript
and for valuable comments, which helped us to improve the paper content.
The work was performed as part of the SAO RAS government contract approved by the
Ministry of Science and Higher Education of the Russian Federation.
Observations at the 6-m telescope BTA are supported by funding from the
Ministry of Science and Higher Education of the Russian Federation. (agreement
No~14.619.21.0004, project identification RFMEFI61914X0004).
EES acknowledges partial support from M.V.Lomonosov Moscow State University Program of Development.
We acknowledge the use of the SDSS database.
Funding for the Sloan Digital Sky Survey (SDSS) has been provided by the
Alfred P. Sloan Foundation, the Participating Institutions, the National
Aeronautics and Space Administration, the National Science Foundation,
the U.S. Department of Energy, the Japanese Monbukagakusho, and the Max
Planck Society. The SDSS Web site is http://www.sdss.org/.
The SDSS is managed by the Astrophysical Research Consortium (ARC) for the
Participating Institutions.
This research is based on observations made with the NASA/ESA Hubble Space Telescope
obtained from the Space Telescope Science Institute, which is operated by the Association of Universities
for Research in Astronomy, Inc., under NASA contract NAS 5-26555. These observations are associated with
program ID GO 11578 (PI A.Aloisi). 

\section{Data availability}

All discussed photometric data for Knots~1--6 are contained in Tables~A1 -- A6 in the Supplementary material
in the electronic form.

\bsp

\clearpage

\appendix

\section{Supplementary materials}
\label{sec:supp}

In Tables~\ref{tab:Knot1} -- \ref{tab:Knot6} we present all results of integrated photometry in B, V and R-bands
for the DDO68 star-forming regions Knots~1--6, located in the so-called 'Northern Ring' or nearby (Knot~6).
The presented magnitudes include the upward corrections of the measured light which account the loss of the light
in the fixed-size aperture for different seeing, as explained in the end of Sect.~\ref{ssec:errors}. The date
in the first column is given in the formate YYYYMMDD. Abbreviations in column Ref. Notes are as follows:
CA3.5 - 3.5m telescope of Calar Alto Observatory, Spain; INT - Isaac Newton Telescope at Canary Icelands, NOT -
Nordic Optical Telescope
at Canary Icelands, KeckII - Keck~II 10m telescope at Mauna Kea, Hawaii; KPNO 4m - Kitt Peak National Observatory
4m telescope, Arizona, USA; VATT - Vatican 1.8m telescope at Mt.Graham International Observatory, Arizona, USA; SDSS -
Apache Point Observatory 2.5m telescope, New Mexico, USA; SAO BTA - the 6m telescope of Special Astrophysical Observatory,
Russia;
KP2.1 - Kitt Peak National Observatory 2.1m telescope, Arizona, USA. V-band magnitudes were adopted from g-filter magnitude
and the colour V--g, measured in these Knots for the similar brightness; HST - 2.6m Hubble Space Telescope;
SAO 1m - 1m telescope of Special Astrophysical Observatory, Russia;  CMO 2.5m - 2.5m telescope of Caucasian Mountain
Observatory of Moscow State University, Russia.

\begin{table*}
\caption{Summary of our $B,V,R$ total magnitude estimates
for DDO68 star-forming region Knot~1
\label{tab:Knot1}
}
\begin{tabular}{lllllrll}
\hline
\multicolumn{1}{l}{Date}  &
\multicolumn{1}{c}{$B$} &
\multicolumn{1}{c}{$\sigma_{\rm B}$} &
\multicolumn{1}{c}{$V$} &
\multicolumn{1}{c}{$\sigma_{\rm V}$} &
\multicolumn{1}{c}{$R$} &
\multicolumn{1}{c}{$\sigma_{\rm R}$} &
\multicolumn{1}{l}{Ref. Notes}     \\
 \hline \\[-0.2cm]
19880214  &18.56 & 0.011 & --    & --    & 18.29 & 0.011 & CA3.5                    \\ 
19940502  &--    & --    & --    & --    & 18.29 & 0.019 & INT 2.54m                \\ %
19950207  &18.57 & 0.024 & 18.44 & 0.019 & --    & --    & NOT 2.56m                \\ 
19971223  &--    & --    & 18.44 & 0.007 & --    & --    & KeckII                   \\ %
19980217  &18.63 & 0.014 & 18.42 & 0.018 & --    & --    & KPNO 4m                  \\ 
20000503  &18.60 & 0.014 & 18.40 & 0.014 & 18.33 & 0.019 & VATT 1.8m                \\ 
20040416  &18.62 & 0.031 & 18.42 & 0.052 & 18.33 & 0.071 & SDSS                     \\  
20050112  &--    & --    & 18.46 & 0.009 & 18.29 & 0.011 & SAO BTA                  \\  
20070209  &--    & --    & 18.44 & 0.030 & --    & --    & KP2.1 interp. from $V-g$ \\ 
20090121  &--    & --    & 18.46 & 0.007 & --    & --    & SAO BTA                  \\
20100502  & --   & --    & 18.38 & 0.080 & --    & --    & HST                      \\ 
20110305  &18.54 & 0.017 & 18.40 & 0.021 & 18.28 & 0.024 & KPNO 0.9m                \\ 
20130217  &18.56 & 0.020 & 18.43 & 0.025 & 18.33 & 0.023 & KPNO 0.9m                \\ 
20150114  &18.58 & 0.009 & 18.45 & 0.008 & 18.31 & 0.007 & SAO BTA                  \\
20160307  &--    & --    & 18.39 & 0.008 & --    & --    & SAO 1m                   \\ 
20160308  &18.62 & 0.013 & 18.40 & 0.009 & 18.28 & 0.015 & SAO 1m                   \\ 
20160407  &--    & --    & 18.37 & 0.008 & --    & --    & SAO 1m                   \\ 
20160517  &--    & --    & 18.37 & 0.020 & --    & --    & SAO 1m                   \\ 
20161022  &--    & --    & 18.37 & 0.018 & --    & --    & SAO 1m                   \\ 
20161124  &18.61 & 0.010 & 18.39 & 0.009 & 18.29 & 0.015 & SAO 1m                   \\ 
20161224  &--    & --    & 18.42 & 0.010 & --    & --    & SAO 1m                   \\ 
20161231  &--    & --    & 18.40 & 0.007 & --    & --    & SAO 1m                   \\ 
20170418  &--    & --    & 18.41 & 0.006 & --    & --    & CMO 2.5m                 \\ 
20170529  &--    & --    & 18.40 & 0.021 & --    & --    & SAO 1m                   \\ 
20171116  &18.55 & 0.010 & 18.41 & 0.009 & 18.31 & 0.009 & SAO BTA                  \\ 
20180219  &--    & --    & 18.40 & 0.006 & --    & --    & CMO 2.5m                 \\ 
20180405  &18.60 & 0.011 & 18.38 & 0.018 & 18.30 & 0.011 & SAO 1m                   \\ 
20180430  &--    & --    & 18.41 & 0.025 & 18.32 & 0.024 & SAO 1m                   \\ 
20181011  &--    & --    & 18.40 & 0.029 & --    & --    & SAO 1m                   \\ 
20190118  &--    & --    & --    & --    & 18.30 & 0.016 & CMO 2.5m                 \\ 
20190203  &--    & --    & 18.39 & 0.014 & --    & --    & SAO 1m                   \\ 
20191026  &18.60 & 0.008 & 18.43 & 0.009 & 18.32 & 0.009 & SAO BTA                  \\ 
20191125  &--    & --    & 18.37 & 0.017 & --    & --    & SAO 1m                   \\ 
20200119  &--    & --    & --    & --    & 18.31 & 0.006 & SAO BTA                  \\ 
20200120  &18.58 & 0.006 & 18.42 & 0.005 & --    & --    & SAO BTA                  \\ 
20200304  &--    & --    & 18.40 & 0.017 & 18.29 & 0.015 & SAO 1m                   \\ 
20200426  &--    & --    & --    & --    & 18.31 & 0.009 & SAO BTA                  \\ 
20201111  &--    & --    & 18.42 & 0.007 & 18.28 & 0.006 & SAO BTA                  \\ 
20210514  &--    & --    & 18.39 & 0.011 & --    & --    & SAO 1m                   \\ 
20210516  &--    & --    & --    & --    & 18.30 & 0.012 & SAO 1m                   \\ 
20211202  &--    & --    & 18.41 & 0.006 & 18.31 & 0.009 & SAO BTA                  \\ 
20220426  &--    & --    & 18.40 & 0.014 & --    & --    & SAO 1m                   \\ 
20221025  &--    & --    & 18.43 & 0.011 & 18.30 & 0.011 & SAO 1m                   \\ 
20221220  &18.55 & 0.005 & 18.43 & 0.008 & 18.28 & 0.005 & SAO BTA                  \\ 
20221225  &18.57 & 0.008 & 18.41 & 0.008 & 18.29 & 0.009 & SAO 1m                   \\ 
20230123  &18.62 & 0.005 & 18.41 & 0.006 & 18.33 & 0.007 & SAO 1m                   \\ 
20230318  &--    & --    & 18.39 & 0.006 & --    & --    & SAO 1m                   \\ 
20231012  &--    & --    & 18.43 & 0.011 & --    & --    & SAO 1m                   \\ 
20231022  &18.56 & 0.008 & 18.41 & 0.009 & --    & --    & SAO BTA                  \\ 

\\[-0.25cm] \hline \\[-0.2cm]
\end{tabular}
\end{table*}


\begin{table*}
\caption{Summary of our $B,V,R$ total magnitude estimates
for DDO68 star-forming region Knot~2
\label{tab:Knot2}
}
\begin{tabular}{lllllrll}
\hline
\multicolumn{1}{l}{Date}  &
\multicolumn{1}{c}{$B$} &
\multicolumn{1}{c}{$\sigma_{\rm B}$} &
\multicolumn{1}{c}{$V$} &
\multicolumn{1}{c}{$\sigma_{\rm V}$} &
\multicolumn{1}{c}{$R$} &
\multicolumn{1}{c}{$\sigma_{\rm R}$} &
\multicolumn{1}{l}{Ref. Notes}     \\
 \hline \\[-0.2cm]                     
19880214  & 19.74 & 0.018 & --    & --    & 19.38 & 0.022 & CA3.5                    \\ 
19940502  & --    & --    & --    & --    & 19.36 & 0.021 & INT 2.54m                \\ %
19950207  & 19.75 & 0.029 & 19.63 & 0.027 & --    & --    & NOT 2.56m                \\ 
19971223  & --    & --    & 19.66 & 0.010 & --    & --    & KeckII                   \\ %
19980217  & 19.76 & 0.028 & 19.62 & 0.039 & --    & --    & KPNO 4m                  \\ 
20000503  & 19.77 & 0.024 & 19.65 & 0.028 & 19.41 & 0.036 & VATT 1.8m                \\ 
20040416  & 19.82 & 0.062 & 19.54 & 0.070 & 19.45 & 0.092 & SDSS                     \\  
20050112  & --    & --    & 19.67 & 0.014 & 19.43 & 0.013 & SAO BTA                  \\  
20070209  & --    & --    & 19.63 & 0.037 & --    & --    & KP2.1 interp. from $V-g$ \\ 
20090121  & --    & --    & 19.65 & 0.010 & --    & --    & SAO BTA                  \\
20100502  & --    & --    & 19.56 & 0.077 & --    & --    & HST                      \\ 
20110305  & 19.73 & 0.027 & 19.61 & 0.039 & 19.42 & 0.046 & KPNO 0.9m                \\ 
20130217  & 19.71 & 0.036 & 19.61 & 0.051 & 19.38 & 0.043 & KPNO 0.9m                \\ 
20150114  & 19.78 & 0.012 & 19.65 & 0.016 & 19.43 & 0.012 & SAO BTA                  \\
20160307  & --    & --    & 19.60 & 0.014 & --    & --    & SAO 1m                   \\ 
20160308  & 19.74 & 0.018 & 19.61 & 0.018 & 19.40 & 0.024 & SAO 1m                   \\ 
20160407  & --    & --    & 19.58 & 0.015 & --    & --    & SAO 1m                   \\ 
20160517  & --    & --    & 19.56 & 0.045 & --    & --    & SAO 1m                   \\ 
20161022  & --    & --    & 19.60 & 0.032 & --    & --    & SAO 1m                   \\ 
20161124  & 19.78 & 0.014 & 19.58 & 0.016 & 19.42 & 0.025 & SAO 1m                   \\ 
20161224  & --    & --    & 19.56 & 0.018 & --    & --    & SAO 1m                   \\ 
20161231  & --    & --    & 19.58 & 0.011 & --    & --    & SAO 1m                   \\ 
20170418  & --    & --    & 19.57 & 0.009 & --    & --    & CMO 2.5m                 \\ 
20170529  & --    & --    & 19.67 & 0.051 & --    & --    & SAO 1m                   \\ 
20171116  & 19.74 & 0.011 & 19.66 & 0.014 & 19.42 & 0.011 & SAO BTA                  \\ 
20180219  & --    & --    & 19.61 & 0.009 & --    & --    & CMO 2.5m                 \\ 
20180405  & 19.73 & 0.018 & 19.58 & 0.037 & 19.38 & 0.022 & SAO 1m                   \\ 
20180430  & --    & --    & 19.61 & 0.056 & 19.48 & 0.054 & SAO 1m                   \\ 
20181011  & --    & --    & 19.63 & 0.068 & --    & --    & SAO 1m                   \\ 
20190118  & --    & --    & --    & --    & 19.39 & 0.022 & CMO 2.5m                 \\ 
20190203  & --    & --    & 19.58 & 0.018 & --    & --    & SAO 1m                   \\ 
20191026  & 19.77 & 0.009 & 19.64 & 0.012 & 19.47 & 0.011 & SAO BTA                  \\ 
20191125  & --    & --    & 19.58 & 0.021 & --    & --    & SAO 1m                   \\ 
20200119  & --    & --    & --    & --    & 19.46 & 0.008 & SAO BTA                  \\ 
20200120  & 19.75 & 0.008 & 19.67 & 0.009 & --    & --    & SAO BTA                  \\ 
20200304  & --    & --    & 19.65 & 0.029 & 19.44 & 0.03  & SAO 1m                   \\ 
20200426  & --    & --    & --    & --    & 19.46 & 0.012 & SAO BTA                  \\ 
20201111  & --    & --    & 19.69 & 0.018 & 19.47 & 0.011 & SAO BTA                  \\ 
20210514  & --    & --    & 19.60 & 0.017 & --    & --    & SAO 1m                   \\ 
20210516  & --    & --    & --    & --    & 19.43 & 0.025 & SAO 1m                   \\ 
20211202  & --    & --    & 19.67 & 0.013 & 19.45 & 0.014 & SAO BTA                  \\ 
20220426  & --    & --    & 19.56 & 0.021 & --    & --    & SAO 1m                   \\ 
20221025  & --    & --    & 19.57 & 0.019 & 19.45 & 0.027 & SAO 1m                   \\ 
20221220  & 19.74 & 0.009 & 19.68 & 0.014 & 19.44 & 0.01  & SAO BTA                  \\ 
20221225  & 19.77 & 0.016 & 19.67 & 0.013 & 19.43 & 0.015 & SAO 1m                   \\ 
20230123  & 19.77 & 0.009 & 19.60 & 0.011 & 19.50 & 0.012 & SAO 1m                   \\ 
20230318  & --    & --    & 19.57 & 0.013 & --    & --    & SAO 1m                   \\ 
20231012  & --    & --    & 19.63 & 0.017 & --    & --    & SAO 1m                   \\ 
20231022  & 19.75 & 0.012 & 19.61 & 0.020 & --    & --    & SAO BTA                  \\ 
\\[-0.25cm] \hline \\[-0.2cm]
\end{tabular}
\end{table*}


\begin{table*}
\caption{Summary of our $B,V,R$ total magnitude estimates for DDO68 SF region~3.
\label{tab:photo_Kn3}
}
\begin{tabular}{llllllll}
\hline
\multicolumn{1}{l}{Date}  &
\multicolumn{1}{c}{$B$} &
\multicolumn{1}{c}{$\sigma_{\rm B}$} &
\multicolumn{1}{c}{$V$} &
\multicolumn{1}{c}{$\sigma_{\rm V}$} &
\multicolumn{1}{c}{$R$} &
\multicolumn{1}{c}{$\sigma_{\rm R}$} &
\multicolumn{1}{l}{Ref. Notes}     \\
 \hline \\[-0.2cm]
19880214   & 20.21 & 0.018 & 20.04 & 0.030 & 19.76 & 0.013 & CA 3.5m phot                        \\ 
19940502   & --    &  --   & 19.99 & 0.030 & 19.81 & 0.020 & INT 2.54m phot                      \\ 
19950207   & 20.21 & 0.018 & 20.08 & 0.026 & --    &  --   & NOT 2.56m phot                      \\ 
19971223   & --    &  --   & 20.09 & 0.010 & --    &  --   & KeckII                              \\ 
19980217   & 20.20 & 0.018 & 20.05 & 0.040 & --    &  --   & KPNO 4m                             \\ 
20000503   & 20.19 & 0.022 & 20.13 & 0.026 & 19.86 & 0.034 & VATT 1.8m                           \\ 
20040416   & 20.22 & 0.050 & 20.01 & 0.057 & 19.90 & 0.072 & SDSS phot                           \\ 
20041109   & 20.12 & 0.060 & 20.11 & 0.050 & --    & --    & BTA cv (1)                          \\ 
20050112   & 20.30 & 0.030 & 20.19 & 0.013 & 19.89 & 0.013 & BTA phot+cv (1)                     \\ 
20070209   & --    &  --   & 19.89 & 0.040 & --    &  --   & KPNO 2.1m (1) interpol. from $V-g$  \\ 
20080111   & 19.90 & 0.030 & 19.93 & 0.020 & --    & --    & SAO BTA cv (1)                      \\ %
20080202   &(19.47)&  --   & 19.50 & 0.050 & --    & --    & APO 3.5m cv (1)                     \\ %
20080204   & 19.51 & 0.070 &(19.52)& 0.050 & --    & --    & SAO BTA cv (1)                      \\ 
20080328   &(19.20)& 0.100 & 19.30 & 0.100 & --    & --    & MMT 6.5m cv (1)                    \\ 
20090121   & --    &  --   & 19.55 & 0.008 & --    &  --   & SAO BTA phot+cv (1)                 \\ %
20100502   & --    &  --   & 19.33 & 0.077 & --    & --    & HST phot (1)                        \\ 
20110305   & 19.74 & 0.022 & 19.65 & 0.032 & 19.39 & 0.036 & KPNO 0.9m                           \\ 
20130217   & 20.02 & 0.034 & 19.88 & 0.048 & 19.64 & 0.044 & KPNO 0.9m                           \\ 
20150114   & 20.20 & 0.012 & 20.17 & 0.015 & 19.93 & 0.011 & SAO BTA phot                        \\
20160307   & --    & --    & 19.96 & 0.013 & --    & --    & SAO 1m phot                         \\ 
20160308   & 20.10 & 0.018 & 19.95 & 0.019 & 19.75 & 0.023 & SAO 1m phot                         \\ 
20160407   & --    & --    & 19.95 & 0.015 & --    & --    & SAO 1m phot                         \\ 
20160517   & --    & --    & 19.92 & 0.045 & --    & --    & SAO 1m phot                         \\ 
20161022   & --    & --    & 19.99 & 0.024 & --    & --    & SAO 1m phot                         \\ 
20161124   & 20.13 & 0.013 & 20.00 & 0.016 & 19.84 & 0.024 & SAO 1m phot                         \\ 
20161224   & --    & --    & 19.89 & 0.019 & --    & --    & SAO 1m phot                         \\ 
20161231   & --    & --    & 19.99 & 0.011 & --    & --    & SAO 1m                              \\ 
20170418   & --    & --    & 19.96 & 0.008 & --    & --    & CMO 2.5m phot                       \\ 
20170529   & --    & --    & 19.93 & 0.044 & --    & --    & SAO 1m phot                         \\ 
20171116   & 20.20 & 0.011 & 20.13 & 0.013 & 19.93 & 0.011 & SAO BTA phot                        \\ 
20180219   & --    & --    & 20.03 & 0.009 & --    & --    & CMO 2.5m phot                       \\ 
20180405   & 20.18 & 0.017 & 19.98 & 0.035 & 19.84 & 0.022 & SAO 1m phot                         \\ 
20180430   & --    & --    & 19.96 & 0.052 & 19.86 & 0.050 & SAO 1m phot                         \\ 
20181011   & --    & --    & 20.05 & 0.066 & --    & --    & SAO 1m phot                         \\ 
20190118   & --    & --    & --    & --    & 19.77 & 0.018 & CMO 2.5m phot                       \\ 
20190203   & --    & --    & 20.00 & 0.018 & --    & --    & SAO 1m phot                         \\ 
20191026   & 20.18 & 0.009 & 20.12 & 0.012 & 19.87 & 0.011 & SAO BTA phot                        \\ 
20191125   & --    & --    & 19.99 & 0.021 & --    & --    & SAO 1m phot                         \\ 
20200119   & --    & --    & --    & --    & 19.87 & 0.008 & SAO BTA phot                        \\ 
20200120   & 20.19 & 0.008 & 20.16 & 0.009 & --    & --    & SAO BTA phot                        \\ 
20200304   & --    & --    & 20.15 & 0.028 & 19.88 & 0.030 & SAO 1m phot                         \\ 
20200426   & --    & --    & --    & --    & 19.89 & 0.011 & SAO BTA phot                        \\ 
20201111   & --    & --    & 20.19 & 0.017 & 19.89 & 0.011 & SAO BTA phot                        \\ 
20210514   & --    & --    & 20.03 & 0.018 & --    & --    & SAO 1m phot                         \\ 
20210516   & --    & --    & --    & --    & 19.93 & 0.027 & SAO 1m phot                         \\ 
20211202   & --    & --    & 20.18 & 0.013 & 19.91 & 0.013 & SAO BTA phot                        \\ 
20220426   & --    & --    & 20.00 & 0.021 & --    & --    & SAO 1m phot                         \\ 
20221025   & --    & --    & 19.95 & 0.020 & 19.85 & 0.026 & SAO 1m phot                         \\ 
20221220   & 20.16 & 0.009 & 20.14 & 0.014 & 19.90 & 0.010 & SAO BTA phot                        \\ 
221225   & 20.14 & 0.017 & 19.96 & 0.013 & 19.85 & 0.015 & SAO 1m phot                         \\ 
230123   & 20.12 & 0.009 & 19.98 & 0.011 & 19.83 & 0.012 & SAO 1m phot                         \\ 
230318   & --    & --    & 20.02 & 0.014 & --    & --    & SAO 1m phot                         \\ 
231012   & --    & --    & 20.05 & 0.017 & --    & --    & SAO 1m phot                         \\ 
231022   & 20.14 & 0.012 & 20.04 & 0.018 & --    & --    & SAO BTA                             \\ 
\\[-0.25cm] \hline \\[-0.2cm]
 \multicolumn{8}{l}{(1) adopted from \citet{DDO68LBV}. 'cv' means the estimate from the spectrum convolution } \\
 \multicolumn{8}{l}{with the respective passband. See details in \citet{DDO68LBV}.} \\
\end{tabular}
\end{table*}


\begin{table*}
\caption{Summary of our $B,V,R$ total magnitude estimates
for DDO68 star-forming region Knot~4
\label{tab:Knot4}
}
\begin{tabular}{lllllrll}
\hline
\multicolumn{1}{l}{Date}  &
\multicolumn{1}{c}{$B$} &
\multicolumn{1}{c}{$\sigma_{\rm B}$} &
\multicolumn{1}{c}{$V$} &
\multicolumn{1}{c}{$\sigma_{\rm V}$} &
\multicolumn{1}{c}{$R$} &
\multicolumn{1}{c}{$\sigma_{\rm R}$} &
\multicolumn{1}{l}{Ref. Notes}     \\
 \hline \\[-0.2cm]
19880214  & 18.76 & 0.012 & --    & --    & 18.59 & 0.015 & CA3.5                    \\ 
19940502  & --    & --    & --    & --    & 18.58 & 0.019 & INT 2.54m                \\ %
19950207  & 18.78 & 0.025 & 18.73 & 0.020 & --    & --    & NOT 2.56m                \\ 
19971223  & --    & --    & 18.70 & 0.008 & --    & --    & KeckII                   \\ %
19980217  & 18.79 & 0.017 & 18.70 & 0.023 & --    & --    & KPNO 4m                  \\ 
20000503  & 18.78 & 0.015 & 18.73 & 0.018 & 18.63 & 0.026 & VATT 1.8m                \\ 
20040416  & 18.83 & 0.033 & 18.71 & 0.060 & 18.63 & 0.065 & SDSS                     \\  
20050112  & --    & --    & 18.72 & 0.010 & 18.58 & 0.011 & SAO BTA                  \\  
20070209  & --    & --    & 18.73 & 0.032 & --    & --    & KP2.1 interp. from $V-g$ \\ 
20090121  & --    & --    & 18.73 & 0.008 & --    & --    & SAO BTA                  \\
20100502  & --    & --    & 18.69 & 0.080 & --    & --    & HST                      \\ 
20110305  & 18.79 & 0.019 & 18.74 & 0.027 & 18.60 & 0.032 & KPNO 0.9m                \\ 
20130207  & 18.80 & 0.023 & 18.75 & 0.031 & 18.63 & 0.030 & KPNO 0.9m                \\ 
20150114  & 18.81 & 0.010 & 18.73 & 0.010 & 18.64 & 0.009 & SAO BTA                  \\
20160307  & --    & --    & 18.70 & 0.009 & --    & --    & SAO 1m                   \\ 
20160308  & 18.78 & 0.014 & 18.68 & 0.013 & 18.58 & 0.019 & SAO 1m                   \\ 
20160407  & --    & --    & 18.71 & 0.011 & --    & --    & SAO 1m                   \\ 
20160517  & --    & --    & 18.69 & 0.03  & --    & -     & SAO 1m                   \\ 
20161022  & --    & --    & 18.73 & 0.024 & -     & --    & SAO 1m                   \\ 
20161124  & 18.81 & 0.010 & 18.73 & 0.012 & 18.63 & 0.018 & SAO 1m                   \\ 
20161224  & --    & --    & 18.70 & 0.013 & --    & --    & SAO 1m                   \\ 
20161231  & --    & --    & 18.69 & 0.008 & --    & --    & SAO 1m                   \\ 
20170418  & --    & --    & 18.70 & 0.007 & --    & --    & CMO 2.5m                 \\ 
20170529  & --    & --    & 18.75 & 0.032 & --    & --    & SAO 1m                   \\ 
20171116  & 18.79 & 0.010 & 18.69 & 0.010 & 18.62 & 0.009 & SAO BTA                  \\ 
20180219  & --    & --    & 18.75 & 0.007 & --    & --    & CMO 2.5m                 \\ 
20180405  & 18.80 & 0.013 & 18.68 & 0.023 & 18.62 & 0.016 & SAO 1m                   \\ 
20180430  & --    & --    & 18.74 & 0.037 & 18.61 & 0.035 & SAO 1m                   \\ 
20181011  & --    & --    & 18.73 & 0.041 & --    & --    & SAO 1m                   \\ 
20190118  & --    & --    & --    & --    & 18.61 & 0.017 & CMO 2.5m                 \\ 
20190203  & --    & --    & 18.76 & 0.016 & --    & --    & SAO 1m                   \\ 
20191026  & 18.80 & 0.008 & 18.68 & 0.010 & 18.61 & 0.010 & SAO BTA                  \\ 
20191125  & --    & --    & 18.76 & 0.017 & --    & --    & SAO 1m                   \\ 
20200119  & --    & --    & --    & --    & 18.58 & 0.007 & SAO BTA                  \\ 
20200120  & 18.81 & 0.006 & 18.69 & 0.006 & --    & --    & SAO BTA                  \\ 
20200304  & --    & --    & 18.75 & 0.020 & 18.63 & 0.023 & SAO 1m                   \\ 
20200426  & --    & --    & --    & --    & 18.61 & 0.009 & SAO BTA                  \\ 
20201111  & --    & --    & 18.68 & 0.009 & 18.60 & 0.008 & SAO BTA                  \\ 
20210514  & --    & --    & 18.70 & 0.013 & --    & --    & SAO 1m                   \\ 
20210516  & --    & --    & --    & --    & 18.61 & 0.017 & SAO 1m                   \\ 
20211202  & --    & --    & 18.70 & 0.008 & 18.58 & 0.010 & SAO BTA                  \\ 
20220426  & --    & --    & 18.72 & 0.016 & --    & --    & SAO 1m                   \\ 
20221025  & --    & --    & 18.69 & 0.014 & 18.59 & 0.015 & SAO 1m                   \\ 
20221220  & 18.80 & 0.006 & 18.68 & 0.009 & 18.57 & 0.007 & SAO BTA                  \\ 
20221225  & 18.77 & 0.009 & 18.72 & 0.010 & 18.58 & 0.010 & SAO 1m                   \\ 
20230123  & 18.81 & 0.006 & 18.78 & 0.007 & 18.61 & 0.010 & SAO 1m                   \\ 
20230318  & --    & --    & 18.67 & 0.009 & --    & --    & SAO 1m                   \\ 
20231012  & --    & --    & 18.75 & 0.013 & --    & --    & SAO 1m                   \\ 
20231022  & 18.79 & 0.008 & 18.75 & 0.011 & --    & --    & SAO BTA                  \\ 
\\[-0.25cm] \hline \\[-0.2cm]
\end{tabular}
\end{table*}


\begin{table*}
\caption{Summary of our $B,V,R$ total magnitude estimates
for DDO68 star-forming region Knot~5
\label{tab:Knot5}
}
\begin{tabular}{lllllrll}
\hline
\multicolumn{1}{l}{Date}  &
\multicolumn{1}{c}{$B$} &
\multicolumn{1}{c}{$\sigma_{\rm B}$} &
\multicolumn{1}{c}{$V$} &
\multicolumn{1}{c}{$\sigma_{\rm V}$} &
\multicolumn{1}{c}{$R$} &
\multicolumn{1}{c}{$\sigma_{\rm R}$} &
\multicolumn{1}{l}{Ref. Notes}     \\
 \hline \\[-0.2cm]
19880214  & 19.23 & 0.010 & --    & --    & 18.99 & 0.011 & CA3.5                    \\ 
19940502  & --    & --    & --    & --    & 19.01 & 0.019 & INT 2.54m                \\ %
19950207  & 19.19 & 0.024 & 19.08 & 0.018 & --    & --    & NOT 2.56m                \\ 
19971223  & --    & --    & 19.10 & 0.008 & --    & --    & KeckII                   \\ %
19980217  & 19.18 & 0.014 & 19.16 & 0.017 & --    & --    & KPNO 4m                  \\ 
20000503  & 19.21 & 0.013 & 19.15 & 0.012 & 19.00 & 0.016 & VATT 1.8m                \\ 
20040416  & 19.19 & 0.015 & 19.12 & 0.018 & 19.05 & 0.030 & SDSS                     \\  
20050112  & --    & --    & 19.09 & 0.009 & 18.99 & 0.011 & SAO BTA                  \\  
20070209  & --    & --    & 19.06 & 0.033 & --    & --    & KP2.1 interp. from $V-g$ \\ 
20090121  & --    & --    & 19.10 & 0.007 & --    & --    & SAO BTA                  \\
20100502  & --    & --    & 19.13 & 0.080 & --    &  --   & HST                      \\ 
20110305  & 19.22 & 0.017 & 19.14 & 0.019 & 19.00 & 0.022 & KPNO 0.9m                \\ 
20130217  & 19.19 & 0.020 & 19.15 & 0.026 & 18.98 & 0.023 & KPNO 0.9m                \\ 
20150114  & 19.22 & 0.009 & 19.10 & 0.008 & 19.00 & 0.007 & SAO BTA                  \\
20160307  & --    & --    & 19.09 & 0.007 & --    & --    & SAO 1m                   \\ 
20160308  & 19.20 & 0.013 & 19.08 & 0.009 & 18.98 & 0.015 & SAO 1m                   \\ 
20160407  & --    & --    & 19.09 & 0.008 & --    & --    & SAO 1m                   \\ 
20160517  & --    & --    & 19.11 & 0.020 & --    & --    & SAO 1m                   \\ 
20161022  & --    & --    & 19.10 & 0.017 & --    & --    & SAO 1m                   \\ 
20161124  & 19.22 & 0.009 & 19.09 & 0.008 & 19.03 & 0.012 & SAO 1m                   \\ 
20161224  & --    & --    & 19.12 & 0.011 & --    & --    & SAO 1m                   \\ 
20161231  & --    & --    & 19.11 & 0.007 & --    & --    & SAO 1m                   \\ 
20170418  & --    & --    & 19.12 & 0.006 & --    & --    & CMO 2.5m                 \\ 
20170529  & --    & --    & 19.10 & 0.019 & --    & --    & SAO 1m                   \\ 
20171116  & 19.22 & 0.010 & 19.12 & 0.009 & 19.04 & 0.009 & SAO BTA                  \\ 
20180219  & --    & --    & 19.13 & 0.006 & --    & --    & CMO 2.5m                 \\ 
20180405  & 19.20 & 0.010 & 19.09 & 0.016 & 19.01 & 0.010 & SAO 1m                   \\ 
20180430  & --    & --    & 19.11 & 0.020 & 19.03 & 0.022 & SAO 1m                   \\ 
20181011  & --    & --    & 19.11 & 0.026 & --    & --    & SAO 1m                   \\ 
20190118  & --    & --    & --    & --    & 19.01 & 0.015 & CMO 2.5m                 \\ 
20190203  & --    & --    & 19.09 & 0.014 & --    & --    & SAO 1m                   \\ 
20191026  & 19.19 & 0.008 & 19.11 & 0.009 & 18.99 & 0.009 & SAO BTA                  \\ 
20191125  & --    & --    & 19.08 & 0.016 & --    & --    & SAO 1m                   \\ 
20200119  & --    & --    & --    & --    & 18.99 & 0.006 & SAO BTA                  \\ 
20200120  & 19.20 & 0.006 & 19.08 & 0.004 & --    & --    & SAO BTA                  \\ 
20200304  & --    & --    & 19.11 & 0.015 & 19.03 & 0.013 & SAO 1m                   \\ 
20200426  & --    & --    & --    & --    & 19.00 & 0.009 & SAO BTA                  \\ 
20201111  & --    & --    & 19.11 & 0.005 & 18.99 & 0.005 & SAO BTA                  \\ 
20210514  & --    & --    & 19.10 & 0.011 & --    & --    & SAO 1m                   \\ 
20210516  & --    & --    & --    & --    & 19.05 & 0.011 & SAO 1m                   \\ 
20211202  & --    & --    & 19.09 & 0.005 & 19.00 & 0.009 & SAO BTA                  \\ 
20220426  & --    & --    & 19.10 & 0.014 & --    & --    & SAO 1m                   \\ 
20221025  & --    & --    & 19.11 & 0.011 & 19.05 & 0.010 & SAO 1m                   \\ 
20221220  & 19.20 & 0.005 & 19.10 & 0.008 & 19.01 & 0.005 & SAO BTA                  \\ 
20221225  & 19.20 & 0.008 & 19.11 & 0.008 & 19.04 & 0.008 & SAO 1m                   \\ 
20230123  & 19.19 & 0.005 & 19.12 & 0.006 & 19.03 & 0.007 & SAO 1m                   \\ 
20230318  & --    & --    & 19.10 & 0.007 & --    & --    & SAO 1m                   \\ 
20231012  & --    & --    & 19.12 & 0.011 & --    & --    & SAO 1m                   \\ 
20231022  & 19.20 & 0.008 & 19.11 & 0.008 & --    & --    & SAO BTA                  \\ 
\\[-0.25cm] \hline \\[-0.2cm]
\end{tabular}
\end{table*}


\begin{table*}
\caption{Summary of our $B,V,R$ total magnitude estimates
for DDO68 star-forming region Knot~6
\label{tab:Knot6}
}
\begin{tabular}{lllllrll}
\hline
\multicolumn{1}{l}{Date}  &
\multicolumn{1}{c}{$B$} &
\multicolumn{1}{c}{$\sigma_{\rm B}$} &
\multicolumn{1}{c}{$V$} &
\multicolumn{1}{c}{$\sigma_{\rm V}$} &
\multicolumn{1}{c}{$R$} &
\multicolumn{1}{c}{$\sigma_{\rm R}$} &
\multicolumn{1}{l}{Ref. Notes}     \\
 \hline \\[-0.2cm]
19880214  & 19.70 & 0.014 & --    & --    & 19.23 & 0.016 & CA3.5                    \\ 
19940502  & --    & --    & --    & --    & 19.26 & 0.020 & INT 2.54m                \\ %
19950207  & 19.66 & 0.026 & 19.41 & 0.021 & --    & --    & NOT 2.56m                \\ 
19971223  & --    & --    & 19.36 & 0.008 & --    & --    & KeckII                   \\ %
19980217  & 19.62 & 0.021 & 19.43 & 0.025 & --    & --    & KPNO 4m                  \\ 
20000503  & 19.70 & 0.018 & 19.45 & 0.019 & --    & 0.027 & VATT 1.8m                \\ 
20040416  & 19.67 & 0.042 & 19.44 & 0.064 & 19.33 & 0.072 & SDSS                     \\  
20050112  & --    & --    & 19.45 & 0.010 & 19.25 & 0.012 & SAO BTA                  \\  
20070209  & --    & --    & 19.40 & 0.041 & --    & --    & KP2.1 interp. from $V-g$ \\ 
20090121  & --    & --    & 19.50 & 0.008 & --    & --    & SAO BTA                  \\
20100502  & --    & --    & 19.55 & 0.080 & --    &  --   & HST                      \\ 
20110305  & 19.63 & 0.021 & 19.43 & 0.028 & 19.25 & 0.032 & KPNO 0.9m                \\ 
20130207  & 19.60 & 0.026 & 19.36 & 0.032 & 19.19 & 0.030 & KPNO 0.9m                \\ 
20150114  & 19.68 & 0.010 & 19.39 & 0.010 & 19.21 & 0.008 & SAO BTA                  \\
20160307  & --    & --    & 19.41 & 0.010 & --    & --    & SAO 1m                   \\ 
20160308  & 19.63 & 0.015 & 19.41 & 0.014 & 19.19 & 0.019 & SAO 1m                   \\ 
20160407  & --    & --    & 19.42 & 0.012 & --    & --    & SAO 1m                   \\ 
20160517  & --    & --    & 19.36 & 0.031 & --    & --    & SAO 1m                   \\ 
20161022  & --    & --    & 19.46 & 0.024 & --    & --    & SAO 1m                   \\ 
20161124  & 19.63 & 0.012 & 19.39 & 0.013 & 19.25 & 0.018 & SAO 1m                   \\ 
20161224  & --    & --    & 19.43 & 0.015 & --    & --    & SAO 1m                   \\ 
20161231  & --    & --    & 19.44 & 0.010 & --    & --    & SAO 1m                   \\ 
20170418  & --    & --    & 19.40 & 0.007 & --    & --    & CMO 2.5m                 \\ 
20170529  & --    & --    & 19.47 & 0.032 & --    & --    & SAO 1m                   \\ 
20171116  & 19.67 & 0.010 & 19.42 & 0.010 & 19.33 & 0.009 & SAO BTA                  \\ 
20180219  & --    & --    & 19.40 & 0.008 & --    & --    & CMO 2.5m                 \\ 
20180405  & 19.69 & 0.015 & 19.42 & 0.026 & 19.28 & 0.016 & SAO 1m                   \\ 
20180430  & --    & --    & 19.47 & 0.043 & 19.29 & 0.038 & SAO 1m                   \\ 
20181011  & --    & --    & 19.47 & 0.048 & --    & --    & SAO 1m                   \\ 
20190118  & --    & --    & --    & --    & 19.25 & 0.023 & CMO 2.5m                 \\ 
20190203  & --    & --    & 19.50 & 0.016 & --    & --    & SAO 1m                   \\ 
20191026  & 19.64 & 0.008 & 19.42 & 0.010 & 19.24 & 0.010 & SAO BTA                  \\ 
20191125  & --    & --    & 19.49 & 0.018 & --    & --    & SAO 1m                   \\ 
20200119  & --    & --    & --    & --    & 19.28 & 0.007 & SAO BTA                  \\ 
20200120  & 19.65 & 0.007 & 19.52 & 0.006 & --    & --    & SAO BTA                  \\ 
20200304  & --    & --    & 19.51 & 0.021 & 19.33 &00.022 & SAO 1m                   \\ 
20200426  & --    & --    & --    & --    & 19.26 & 0.010 & SAO BTA                  \\ 
20201111  & --    & --    & 19.46 & 0.010 & 19.25 & 0.008 & SAO BTA                  \\ 
20210514  & --    & --    & 19.49 & 0.014 & --    & --    & SAO 1m                   \\ 
20210516  & --    & --    & --    & --    & 19.27 & 0.018 & SAO 1m                   \\ 
20211202  & --    & --    & 19.41 & 0.008 & 19.25 & 0.010 & SAO BTA                  \\ 
20220426  & --    & --    & 19.49 & 0.017 & --    & --    & SAO 1m                   \\ 
20221025  & --    & --    & 19.42 & 0.015 & 19.30 & 0.018 & SAO 1m                   \\ 
20221220  & 19.62 & 0.007 & 19.46 & 0.010 & 19.30 & 0.008 & SAO BTA                  \\ 
20221225  & 19.61 & 0.012 & 19.39 & 0.011 & 19.25 & 0.011 & SAO 1m                   \\ 
20230123  & 19.66 & 0.007 & 19.49 & 0.008 & 19.27 & 0.010 & SAO 1m                   \\ 
20230318  & --    & --    & 19.49 & 0.011 & --    & --    & SAO 1m                   \\ 
20231012  & --    & --    & 19.47 & 0.014 & --    & --    & SAO 1m                   \\ 
20231022  & 19.63 & 0.010 & 19.44 & 0.011 & --    & --    & SAO BTA                  \\ 
\\[-0.25cm] \hline \\[-0.2cm]
\end{tabular}
\end{table*}

\label{lastpage} 
 
\end{document}